\documentclass[twocolumn,floats,floatfix,amssymb,nofootinbib,prd,superscriptaddress,aps]{revtex4-2}
\usepackage{graphicx,amssymb,amsmath,amsthm,amsfonts,epsfig}
\usepackage[linktocpage,colorlinks=true,citecolor=OliveGreen,linkcolor=Maroon,urlcolor=Maroon]{hyperref}
\usepackage[usenames]{color}
\usepackage{bm}
\usepackage{dcolumn}
\usepackage[utf8]{inputenc}
\usepackage{latexsym}
\usepackage{rotating}
\usepackage{tabularx}
\usepackage{braket}
\usepackage{esvect}
\usepackage{enumerate}
\usepackage{footmisc}
\usepackage{array}
\usepackage{tensor}
\usepackage{mathtools}
\usepackage{url}
\usepackage[dvipsnames]{xcolor}
\usepackage{comment}
\usepackage{multirow}
\usepackage{nicematrix}
\usepackage{graphics}
\usepackage{textcomp}
\usepackage{stmaryrd}
\usepackage[normalem]{ulem}
\usepackage{verbatim}
\usepackage{soul,xcolor}
\usepackage{xspace}
\usepackage{xfrac}
\usepackage{ragged2e}
\setstcolor{red}
\usepackage{esint}
\usepackage{physics}
\usepackage{subcaption}

\def\be{\begin{equation}}
\def\ee{\end{equation}}

\newcommand{\beq}{\begin{eqnarray}}

\newcommand{\eeq}{\end{eqnarray}}

\newcommand{\mD}{\mathcal{D}}
\newcommand{\lr}[1]{\left( #1\right)}

\newcommand{\overbar}[1]{\mkern 1.5mu\overline{\mkern-1.5mu#1\mkern-1.5mu}\mkern 1.5mu}
\newcommand\kZ{\overset{k}{Z}\vphantom{Z}}
\newcommand\kZZ{\overset{1}{Z}\vphantom{Z}}
\newcommand\kC{\overset{k}{\mathcal{C}}\vphantom{\mathcal{C}}}
\newcommand\kA{\overset{k}{\mathcal{A}}\vphantom{\mathcal{A}}}
\def\be{\begin{equation}}
\def\ee{\end{equation}}
\def\bea{\begin{eqnarray}}
\def\eea{\end{eqnarray}}
\def\beq{\begin{eqnarray}}
\def\eeq{\end{eqnarray}}

\begin{document}
\newcommand{\AEI}{\affiliation{Max Planck Institute for Gravitational Physics (Albert Einstein Institute), Am M\"uhlenberg 1, Potsdam 14476, Germany}}

\title{Gravitational waves from b-EMRIs:\\
Doppler shift and beaming, resonant excitation, helicity oscillations and self-lensing}
\author{João S. Santos}
\affiliation{CENTRA, Departamento de F\'{\i}sica, Instituto Superior T\'ecnico -- IST, Universidade de Lisboa -- UL,
Avenida Rovisco Pais 1, 1049-001 Lisboa, Portugal}
\affiliation{Center of Gravity, Niels Bohr Institute, Blegdamsvej 17, 2100 Copenhagen, Denmark}
\author{Vitor Cardoso}
\affiliation{CENTRA, Departamento de F\'{\i}sica, Instituto Superior T\'ecnico -- IST, Universidade de Lisboa -- UL,
Avenida Rovisco Pais 1, 1049-001 Lisboa, Portugal}
\affiliation{Center of Gravity, Niels Bohr Institute, Blegdamsvej 17, 2100 Copenhagen, Denmark}
\author{José Natário}
\affiliation{CAMGSD, Departamento de Matem\'{a}tica, Instituto Superior T\'ecnico -- IST, Universidade de Lisboa -- UL,
Avenida Rovisco Pais 1, 1049-001 Lisboa, Portugal}
\author{Maarten van de Meent}
\affiliation{Center of Gravity, Niels Bohr Institute, Blegdamsvej 17, 2100 Copenhagen, Denmark}
\AEI

\date{\today}

\begin{abstract}
We study gravitational waves from a stellar-mass binary orbiting a spinning supermassive black hole, a system referred to as a binary extreme mass ratio inspiral (b-EMRI). 
%
%
We use Dixon's formalism to describe the stellar-mass binary as a particle with internal structure, and keep terms up to quadrupole order to capture the generation of gravitational waves by the inner motion of the stellar-mass binary. The problem of emission and propagation of waves is treated from first principles using black hole perturbation theory. In the gravitational waveform at future null infinity, we identify for the first time Doppler shifts and beaming due to the motion of the center of mass, as well as helicity breaking gravitational lensing, and resonances with ringdown modes of the supermassive black hole. We establish that previously proposed phenomenological models inadequately capture these effects.
\end{abstract}

\maketitle

\noindent\textbf{\emph{Introduction.}} 
The detection of gravitational waves (GWs) from binary black hole (BH) mergers 
has provided us with priceless information on strong field gravity~\cite{LIGOScientific:2016vbw,KAGRA:2021vkt,LIGOScientific:2020ibl,Barack:2018yly,Cardoso:2019rvt,Berti:2025hly}. Merger rates of stellar-mass BHs are now known to good precision, and a number of key questions in fundamental physics will be addressed with more sensitive detectors and precise measurements, or serendipitous discoveries~\cite{Barack:2018yly,Cardoso:2019rvt,Berti:2025hly}. Open challenging questions remain, namely regarding the impact of the environments in which compact binaries are formed and evolve. Space-based detectors like LISA~\cite{LISA:2017pwj}, DECIGO~\cite{Kawamura:2020pcg} and TianQin~\cite{TianQin:2015yph} will follow the evolution of some binary systems for months or years, and thus obtain crucial information about astrophysical environments~\cite{Barausse:2020rsu,LISA:2022kgy,Barack:2018yly}. 

Supermassive black holes (SMBHs) powering active galactic nuclei are promising engines for the formation of binaries of stellar mass BHs through various dynamical channels~\cite{Addison:2015bpa,Bellovary:2015ifg,Bartos:2016dgn,Stone:2016wzz,Hoang:2017fvh,Tagawa:2019osr}. If the resulting binary is stable against tidal disruption from the SMBH~\cite{Hills_1988,Suzuki:2020vfw} while remaining gravitationally bound to the latter, a hierarchical triple is formed~\cite{Addison:2015bpa,Naoz:2011eic,Naoz:2011mb,Will:2020tri,Chen:2018axp,Peissker:2024ade}. There is tentative evidence suggesting that the transient feature detected by the Zwicky Transient Facility~\cite{Masci:2018neq, Graham:2019qsw, Graham:2020gwr} associated to the event GW190521~\cite{LIGOScientific:2020iuh} is a BH merger occurring in the accretion disk of an active galactic nucleus~\cite{Toubiana:2020drf,Gamba:2021gap,Sberna:2022qbn}.

\begin{figure}[ht!]
\centering
\includegraphics[width=.45 \textwidth]{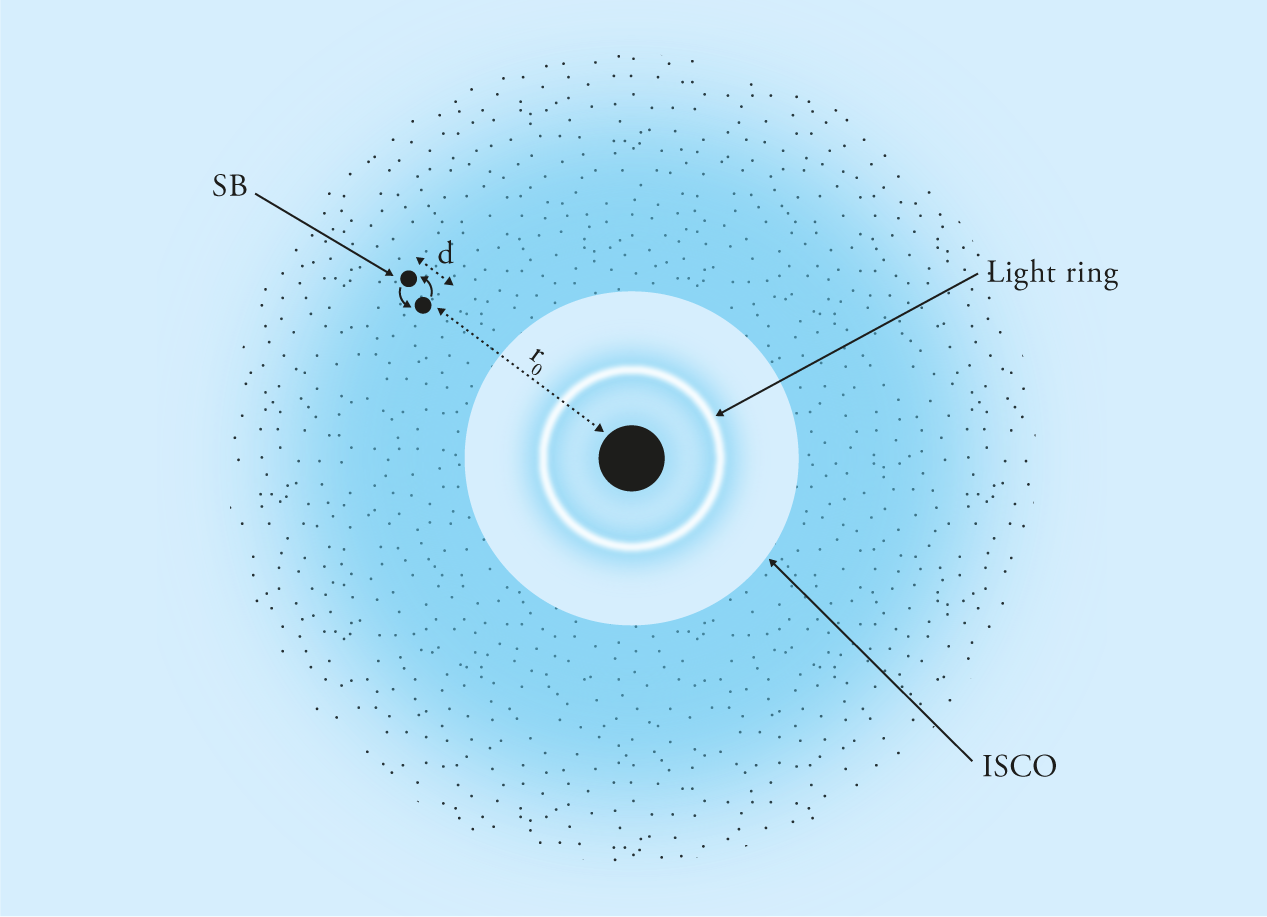}
\caption{\justifying Depiction of the b-EMRI system that we study. A Kerr SMBH sits at the center, and acts as the primary of a b-EMRI system. The ISCO, light ring and horizon of the SMBH all play a role in the GW emission from this system. The secondary binary (SB, not to scale) is represented in the environment of the SMBH. Adapted from Ref.~\cite{Cardoso:2021vjq}. }
\label{fig:b_EMRI}
\end{figure}
The triple systems described above are composed of a secondary stellar mass binary (SB) orbiting the SMBH, see Fig.~\ref{fig:b_EMRI}; given the disparate scales, they are usually referred to as binary extreme mass ratio inspirals (b-EMRIs)~\cite{Chen:2018axp, Chen:2017xbi,Han:2018hby,Cardoso:2021vjq,Meng:2024kug,Jiang:2024mdl,Yin:2024nyz}. GWs from b-EMRIs have been studied by introducing {\it phenomenological} changes to the waveforms of an isolated binary, namely signatures of Doppler shifts~\cite{Cisneros:2012sk,Bonvin:2016qxr,Inayoshi:2017hgw,Robson:2018svj,Tamanini:2019usx,Meiron:2016ipr, Wong:2019hsq,Randall:2018lnh,Han:2018hby,Yan:2023pyo,Zwick:2025wkt,Takatsy:2025bfk,Cusin:2024git}, aberration~\cite{Torres-Orjuela:2018ejx,Torres-Orjuela:2020cly,Torres-Orjuela:2020oxq,Bonvin:2022mkw,Cusin:2024git}, gravitational redshift~\cite{Chen:2017xbi}, and helicity-dependent strong field lensing~\cite{Kubota:2024zkv, Pijnenburg:2024btj,Chan:2025wgz,DOrazio:2019fbq,Oancea:2022szu,Oancea:2023hgu}. Simple models for the SB found that these systems can excite ringdown modes of the SMBH~\cite{Cardoso:2021vjq,Yin:2024nyz}.

Here we develop a formalism to study the generation and propagation of GWs from b-EMRIs from first principles, where all the above physics is borne out naturally, and rigorously: nothing in the waveforms discussed in this work arises from {\it ad hoc} modifications. We find evidence for all the phenomenology previously reported in the literature, but we also find that existing non-relativistic models are unable to satisfactorily capture some features of the b-EMRI waveforms we obtain, which are fully in a strong field regime. 

\noindent\textbf{\emph{Modeling the secondary binary.}} 
%
We take the SMBH geometry to be described by the Kerr family with mass $M$ and spin $a$, see the Supplemental Material (SM), and we take, for simplicity, a SB of two non-spinning and equal-mass $\mu\ll M$ components, separated by a distance $d$.  The SB orbits the SMBH at a (Boyer-Lindquist) radius $r_0\gg d$. 

Leveraging the hierarchy of scales, the SB is described using Dixon's formalism for extended mass distributions~\cite{Dixon:1970zza,Dixon:1974xoz,Dixon:2015vxa,Harte:2011ku,Costa:2012cy, Gralla:2010xg}. This approach amounts to representing the SB as point particle with internal structure encoded in its multipole moments. To capture the radiation produced by the inner motion of the SB components, we keep terms up to quadrupole order. The validity of Dixon's formalism relies on spacetime being approximately flat over the SB, which implies $d\ll\sqrt{r_0^3/M}$. 

The SB is described by its four-momentum $p^\mu$, spin $S^{\mu \nu}$, and quadrupole moment $J^{\mu \nu \rho \sigma}$. These are tensors defined over a reference worldline $z^\mu(\tau)$, where $\tau$ is the proper time along the worldline. In general, the internal multipole structure of the SB affects its trajectory. However, these effects take place over timescales much longer than the orbital timescales around the SMBH, set by $\Omega_0 \sim \sqrt{M/r_0^3}$, and we neglect them: we take $z^\mu(\tau)$ to be a circular geodesic. Within this approximation, $z^\mu(\tau)$ is the center of mass of the SB~\cite{Costa:2014nta}. We don't include backreaction from GW emission, and we neglect effects of the tidal field from the SMBH on the inner orbital dynamics, like the Kozai-Lidov oscillations~\cite{Kozai:1962zz,Lidov:1962wjn,Naoz:2012bx,Yang:2017aht,Cardoso:2021qqu,Maeda:2023uyx,Maeda:2025row,Camilloni:2023rra,Camilloni:2023xvf,Farina:2025fbv,Grilli:2024fds,Cocco:2025adu}.

Within this multipolar expansion framework, the SB can be described by an effective energy momentum tensor with support on the circular geodesic $z^\mu (\tau)$~\cite{Steinhoff:2009tk}:
\begin{align}
&T^{\mu \nu} (x) = \int d\tau \left(\left(u^{(\mu} p ^{\nu)} + \frac{1}{3} R_{\rho \sigma \delta}^{\quad (\mu}J^{\nu) \delta \sigma \rho} \right)\delta_{(4)}\right. \nonumber \\
&\left. - \nabla_\rho \left(S^{\rho (\mu} u^{\nu)} \delta_{(4)} \right)  - \frac{2}{3} \nabla_\sigma \nabla_\rho \left(J^{\sigma(\mu \nu)\rho} \delta_{(4)} \right)\right) \, , \label{eq:canonical_EM_tensor}
\end{align}
where $\delta_{(4)} = \delta_{(4)} (x-z(\tau))/\sqrt{-g}$, with $g$ the determinant of the metric tensor, and $u^\mu = dz^\mu /d \tau$. To obtain the spin and quadrupole moment tensors, we introduce a series of frames that are Fermi-Walker transported along the circular geodesic~\cite{Poisson:2011nh}. We allow for a generic inclination of the SB spin w.r.t the orbital angular momentum and call this angle $\iota_{\rm SB}$. The specific form of the multipole moments and their derivation is given in the SM. The key point is that this procedure introduces two new frequencies in the problem, the precession frequency $\Omega_{\rm P}$\cite{Rindler_Perlick_1990, vandeMeent:2019cam} and the intrinsic frequency of the SB inner motion $\Omega_{\rm SB}$:
\begin{equation}
    \Omega_{\rm P} = \frac{1}{u^t}\sqrt{\frac{M}{r_0^3}} \, , \quad  \Omega_{\rm SB} = \frac{1}{u^t}\sqrt{\frac{2 \mu}{d^3}} \, . \label{eq:frequencies_BL}
\end{equation}
Note that the SB is stable against tidal disruption due to the SMBH if it satisfies the Hills criterion
\cite{Hills_1988, Cardoso:2021vjq, Addison:2015bpa, Suzuki:2020vfw}, $d < R_{\rm Hills} \sim (2 \mu / M)^{1/3} r_0 \, $. In Ref.~\cite{Suzuki:2020vfw}, it was shown that relativistic effects tighten this upper bound by up to a factor 4, and so we take $d<R_{\rm Hills}/4$ (see also Ref.~\cite{Yin:2024nyz}). It is worth pointing out that this criterion implies a separation of the outer and inner orbital timescales of the SB in the form
\begin{equation}
    \left(\frac{\Omega_{\rm SB}}{\Omega_0}\right)^2 \sim \left(\frac{R_{\rm{Hills}}}{d}\right)^3 > 64 \, . 
    \label{eq:hierarchy_omega}
\end{equation}
%
%
\noindent\textbf{\emph{Gravitational wave generation.}} 
%
Since $\mu \ll M$, we are in the realm of BH perturbation theory, and we use the Teukolsky formalism for GW generation (details in the SM). The Teukolsky equation describes the radiative degrees of freedom of a spin-$s$ massless field~\cite{Teukolsky:1973ha,Teukolsky:1974yv}, encapsulated in a master variable $\psi$ sourced by matter fields $\mathcal{T}$. Since we are interested in extracting the GWs at future null infinity, the master variable we solve for is the Weyl scalar $\psi_4$~\cite{Newman:1961qr,Detweiler:1978ge}. Asymptotically, $\psi_4$ takes the form
\begin{equation}
    \psi_4 \sim \sum_{\substack{\ell, m, p ,q}} {Z}^\infty _{\ell m p q} \, _{-2}S^{} _{ \ell m \omega_{mpq}} (\theta) e^{i m \phi} \frac{e^{-i\omega_{mpq} u}}{r}  \, , \label{eq:psi4}
\end{equation}
where $\{u,r,\theta,\phi \}$ are (retarded) Boyer-Lindquist coordinates in the SMBH spacetime,  and $ _{-2}S^{} _{ \ell m \omega}$  are the $s=-2$ spin-weighted spheroidal harmonics~\cite{Press:1973zz,1986JMP....27.1238L} with spheroidicity $a\omega$, angular mode number $\ell\geq2$ and azimuthal mode number $m$ in the range $-\ell \leq m \leq \ell$. The frequencies excited by the system are 
\begin{equation}
    \omega_{mpq} = m \,  \Omega_0 + p \, \Omega_{\rm P} + q \, \rm \Omega_{\rm SB} \, , \label{eq:frequencies}
\end{equation}
where $p\in\{0,\pm1,\pm2\}$ and $q\in\{0,\pm2\}$ are the precession and quadrupole mode numbers. Given the hierarchy of scales in Eq.~\eqref{eq:hierarchy_omega}, these frequencies naturally separate into families with low ($q=0$) and high ($q=\pm 2$) frequency. While the low-frequency modes are excited by all the terms in the energy momentum tensor in Eq.~\eqref{eq:canonical_EM_tensor}, only the terms involving the quadrupole tensor excite high-frequency modes. The amplitudes ${Z}^\infty _{\ell m p q}$ take the form
\begin{equation}
    Z_{\ell m p q}^\infty = \sum_ 
    {i= 0} ^{4}\ \sum_ 
    {j = 0} ^{4-i} \mathcal{A}^{(i,j)}_{\ell m p q} \frac{d^i }{dr^i} R^{H}\left(r_0 \right) \ \frac{d^j}{d\theta^j}  \bar S \left( \pi/2\right) \, , \label{eq:amplitudes}
\end{equation} 
where $\mathcal{A}$ depends only on the outer and inner orbital parameters of the SB, $R^{H}\equiv R^{H}_{\ell m \omega_{mpq}}$ is a solution to the homogeneous Teukolsky equation satisfying purely ingoing boundary conditions at the horizon, and $\bar S \equiv {_{-2}\bar{S}_{\ell m \omega_{mpq}}}$. Finally, a simple equation relates the Weyl scalar to the strain at future null infinity:
\begin{equation}
    \psi_4 \sim \frac{1}{2} \partial_t^2 (h^{TT}_+ - i h^{TT}_\times) \, .
    \label{eq:strain}
\end{equation}

To obtain the waveform generated by the b-EMRI, we must then calculate the amplitudes ${Z}^\infty _{\ell m p q}$. This procedure is described in more detail in the SM. The quantities $\mathcal{A}$ in Eq~\eqref{eq:amplitudes} are obtained analytically, while the eigenfunctions $R^H$ and $\bar S$ are calculated semi-analytically~\cite{Mano:1996vt,Fujita:2004rb,Leaver:1985ax,Berti:2005gp,vandeMeent:2015lxa}. The signal is a sum over harmonics $\ell$ in Eq.~\eqref{eq:psi4} that must converge to some prescribed accuracy. This is a key issue, sometimes overlooked: the further the SB is placed from the center of coordinates, the larger the number of harmonics needed~\cite{Bonetti:2017hnb}. For each value of $p$ and $q$ we sum up to $\ell = \ell_{\rm{max}}$, the value for which a convergence criterion is satisfied (roughly, that the amplitude of the $\ell_{\rm{max}}$ mode is $10^5$ smaller than the peak value, see the SM). This is most important for the high-frequency modes, for which we found $\ell_{\rm max}\sim \Omega_{\rm SB} \, r_0$, in agreement Ref.~\cite{Boyle:2015nqa}. We find that for $\ell \gtrsim \ell_{\rm max}$ the amplitudes exhibit exponential convergence $ Z_{\ell m p q}^\infty \sim 10^{-\beta \ell}$ for $\beta \gtrsim 0.2$. With our formalism, we recover well known results in the EMRI and Newtonian limits (see SM).

%
\noindent\textbf{\emph{Results.}} 
%
%
We now focus on a fiducial set of parameters, corresponding to a SB on a prograde circular geodesic with $\iota_{\rm SB}=0$ (intrinsic SB spin aligned with orbital angular momentum). In this geometric configuration, the low-frequency part of the signal contains only modes with $p=q=0$, while the high-frequency part contains modes with $p=-q=\pm2$. The parameters used in our fiducial simulations are shown in Table~\ref{tab:parameters_fiducial}. This system can represent a SB of two BHs with mass $\mu = 10 M_\odot$ each, separated by $d=500\mu$, well within Newtonian dynamics, orbiting a SMBH with $M = 10^5 M_\odot$. The low-frequency signal of this system is a simple circular EMRI waveform, amply studied in the literature~\cite{Drasco:2005kz}, so we focus on the high-frequency content, $\omega\sim0.1 \, {\rm Hz}$.
This system is stable against tidal disruption according to Hills criterion~\eqref{eq:hierarchy_omega}. 
\begin{table}[]
    \centering
    \begingroup
\setlength{\tabcolsep}{6pt} 
\renewcommand{\arraystretch}{1.2} 
\begin{tabular}{l|c|c}
    \textbf{Quantity} & \textbf{Symbol} & \textbf{Value} \\
    \hline
        Orbital radius & $r_0/M$ & 10 \\
        Spin of SMBH & $a/M$ & 0.7 \\
        Inclination & $\iota_{\rm SB}$ & 0 \\
        SB component mass & $\mu /M $ & $10^{-4}$ \\
        SB separation & $d /\mu$ & $500$ \\
        Orbital frequency  & $M \Omega_{0} $ & 3.09 $\times10^{-2}$ \\
        Precession frequency  & $M {\rm {\rm \Omega_P}} $ & 2.67$\times10^{-2}$ \\
        Frequency of SB  & $M {\rm \Omega_{SB}} $ & 1.07 \\
        \hline
    \end{tabular}
\endgroup
    \caption{\justifying Parameters for the fiducial simulation: a spin-aligned b-EMRI system in a prograde circular geodesic around a Kerr SMBH of mass $M$.}
    \label{tab:parameters_fiducial}
\end{table}
\begin{figure}[ht!]
    \centering
    \includegraphics[width=0.95\linewidth]{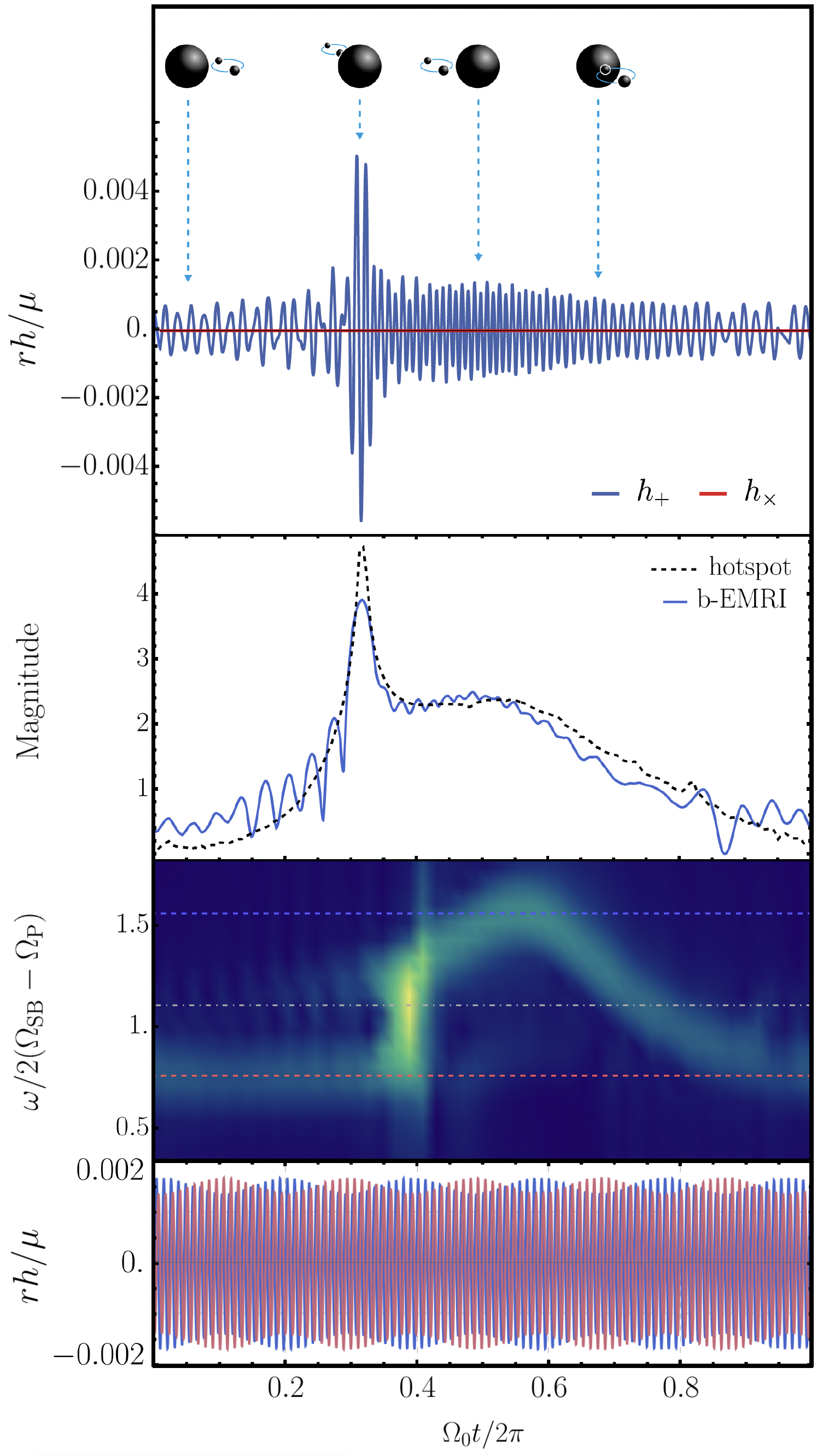}
    \caption{ \justifying Results for GW signal from the fiducial system in Table~\ref{tab:parameters_fiducial}; horizontal axis spans one orbit of the SB around the SMBH. From top to bottom: \emph{First:} waveform for an edge-on observer $\theta=\pi/2$, $\phi=0$, with sketches indicating the SB position in the image plane. \emph{Second:} ``instantaneous" magnitude for the first signal, compared with the results for a hotspot orbiting a Kerr BH with the same parameters. \emph{Third:} spectrogram of first signal showing absolute value of the Fourier transform of $h_+$ in the $t-\omega$ plane. Blue and red lines are predictions for the maximum and minimum frequencies observed as a result of Doppler shifts. Gray dot-dashed line indicates the frequency of the $\ell=m=8$, $n=0$ quasi-normal mode of the SMBH. \emph{Fourth:} waveform for a face-on observer $\theta=0$, $\phi=0$. The analysis is performed in the text.
    }
    \label{fig:waveform}
\end{figure}

Figure~\ref{fig:waveform} summarizes our main results, with different information on the high-frequency signal produced by the b-EMRI. Note that these are all first-principles results, with no phenomenological artifacts added. The horizontal axis spans one orbit of the SB around the SMBH. The top panel shows the waveform seen by an edge-on observer at $\theta=\pi/2$, $\phi=0$. The cross polarization vanishes because the inner orbit of the SB is contained in the equatorial plane; for generic values of $\iota_{\rm SB}$ this is no longer the case. Even by eye, it is easy to identify the signatures of Doppler modulation and gravitational lensing.

\noindent \emph{GW lensing.} The second (from top) panel of Fig.~\ref{fig:waveform} shows the magnitude of the GW radiation, as defined in the SM. We compare our result for the instantaneous magnitude with the magnitude for an isotropically-emitting hotspot in the same orbit around the same Kerr BH~\cite{Hamaus:2008yw,Rosa:2022toh}, using the backwards ray tracing code GYOTO~\cite{Vincent:2011wz}. The curves agree remarkably well: they exhibit the same features, and the peak reflects the appearance of an Einstein ring as the SB passes behind the SMBH. Irregularities in the b-EMRI curve and the broadness of the peak are due to the averaging procedure described in the SM. Using the information from ray tracing, we identify the position of the b-EMRI in the image plane, relative to the SMBH, at different points in the waveform (sketches in top panel of Fig.~\ref{fig:waveform}).

\noindent \emph{Doppler modulation and beaming.} The third panel of Fig.~\ref{fig:waveform} shows the spectrogram of the waveform. The frequency $\omega$ is normalized to $ 2(\Omega_{\rm SB}-\Omega_{P})$, as this becomes the intrinsic frequency of the GWs emitted by the SB as seen by an asymptotic observer, for $\iota_{\rm SB}=0$ (see the SM). Here we quantitatively identify the signatures of Doppler modulation -- the frequency oscillates over time -- and beaming -- the intensity is higher when the frequency signal is blueshifted. The blue and red dashed lines indicate theoretical predictions for the maximum blueshift and redshift, respectively, as obtained using the model derived in the SM, 
\begin{equation}
    \omega_{\pm} = \Omega_{\rm SB} \left(1 + \Omega_0 \frac{u_\phi \pm (e_\phi)_\phi}{u_t \pm (e_\phi)_t}\right)^{-1}, \label{eq:Doppler}
\end{equation}
where $\omega_{\pm}$ are the maximally blueshifted and redshifted frequencies, and $(e_\phi)^\mu$ is a unit spacelike four-vector, with components $((e_\phi)^t,0,0,(e_\phi)^\phi)$ in Boyer-Lindquist coordinates, and orthogonal to $u^\mu$. The predictions are in excellent agreement with the numerical results.

\noindent \emph{QNM resonances.} High frequency GWs can resonantly excite the quasinormal mode (QNM) frequencies of the SMBH~\cite{Berti:2009kk,Berti:2025hly}, $M \omega_{mpq}  \sim M \omega^{\rm QNM}_{\ell m n}\sim \mathcal{O}(1)$, with $n$ an overtone number. This occurs when $\omega_{mpq}\approx {\rm Re}\left[\omega^{\rm QNM}_{\ell m n} \right] $, generating a peak in the amplitude $Z_{\ell m p q}^\infty$. Using analytical expressions for QNMs in the eikonal limit $\ell \gg 1$~\cite{Cardoso:2008bp,Berti:2009kk,Yang:2012he,Dolan:2010wr,Berti:2025hly}, we expect the $m=\ell$, $n=0$ mode (the longest lived) to be excited for 
\begin{equation}
    \ell = m \sim b \, \omega_{mpq} \sim b \, q \, \Omega_{\rm SB }\, ,
    \label{eq:resonance_estimator}
\end{equation}
with $b\sim1$ given in the SM. For our fiducial simulation, we expect the $\ell=m=8$ QNM to be resonantly excited,
\beq
    &M \omega^{\rm QNM}_{(8,8,0)}& \approx 2.30 - 0.0861 \, i \, .
\eeq
A gray dot-dashed line in the third panel of Fig.~\ref{fig:waveform} indicates the real part of the frequency of this QNM. We find strong evidence that this SMBH mode is indeed resonantly excited (cf. the SM); a more detailed study is in preparation.

\noindent \emph{Helicity dependent scattering.} The bottom panel of Fig.~\ref{fig:waveform} shows the waveform for a face-on observer, $\theta=0$, $\phi=0$. Even in this simple case, there is an observable amplitude modulation with frequency $4 \Omega_0$. This is due to helicity-dependent GW lensing, a well known result in wave optics~\cite{Dolan:2008kf,Pijnenburg:2024btj,Chan:2025wgz}. The total waveform in this case is the combination of a transmitted component, which does not interact with the SMBH, and a scattered component, $h = h_{\rm T} + h_{\rm S}$. Varying the orbital radius $r_0$, we find that the ratio of the two follows the rule    
\begin{equation}
    |h_{\rm S}|/ |h_{\rm T}| \approx M /r_0 \, . \label{eq:helicity}
\end{equation}
This is just a factor two off the prediction obtained in Ref.~\cite{Pijnenburg:2024btj}, which may well be explained by their use of a low-frequency expansion (certainly not valid for our system) in combination with a framework where a plane wave hits the SMBH.

\noindent \emph{Phenomenological models.}
%
%
\begin{figure}[t]
\centering
\includegraphics[width=.45 \textwidth]{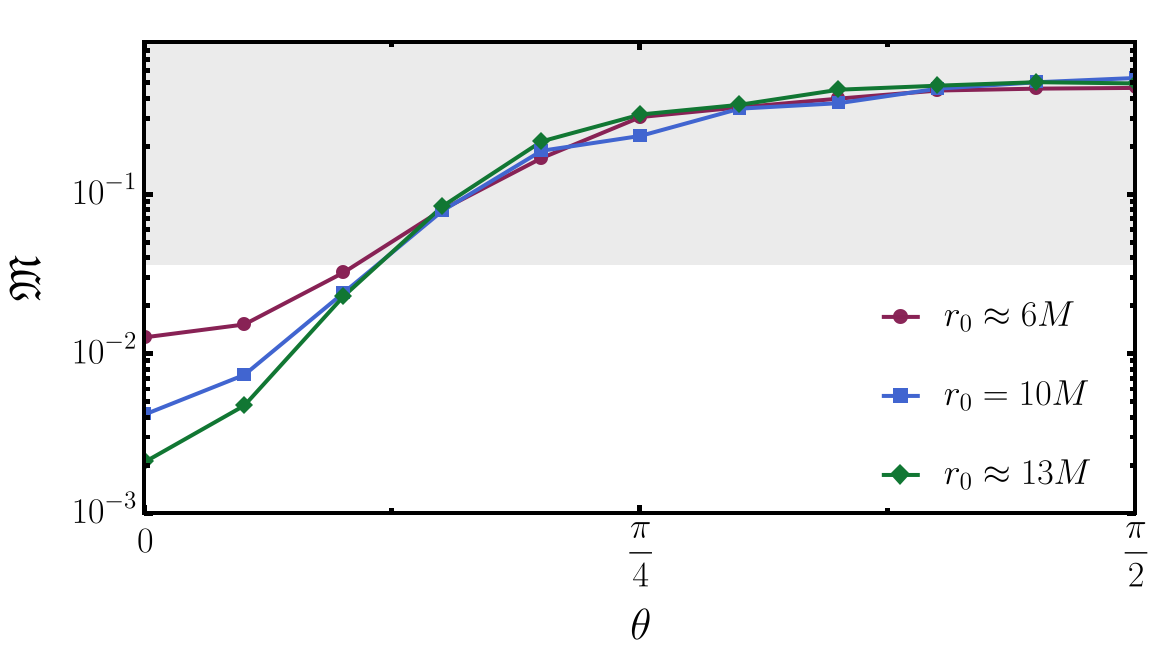}
\caption{\justifying Mismatch curves between waveforms obtained with our b-EMRI model and waveforms built with a non-relativistic Doppler modulation added to the signal from an isolated SB. The mismatch $\mathfrak{M}$ is obtained by maximizing the overlap between the waveforms over the initial phase $\varphi_0$, initial time $t_0$, line of sight velocity $v_{\rm los}$ and frequency $f_{\rm gw}$ (see SM). We vary the observing angle in the x-axis. Different curves correspond to placing the SB at different orbital radii. 
}
\label{fig:mismatch}
\end{figure}
Previous studies of b-EMRIs added a phenomenological, non-relativistic model for the Doppler shift induced by the motion of the SB around the SMBH, neglecting all other strong field effects that we have just discussed~\cite{Bonvin:2016qxr,Inayoshi:2017hgw,Robson:2018svj,Tamanini:2019usx,Meiron:2016ipr, Wong:2019hsq,Randall:2018lnh,Han:2018hby,Yan:2023pyo}. Given the complex features in our waveform, we don't expect these models to capture b-EMRI waveforms in the strong field regime. One can quantify this property by computing the mismatch~\cite{Owen:1995tm,Apostolatos:1995pj,Flanagan:1997kp} (details in the SM) between our waveforms and the non-relativistic model, for various observing angles. We minimize the mismatch over all parameters of the waveform model except the observing angle to get $\mathfrak{M}$. This quantity measures how effective the ``added-on nonrelativistic Doppler'' template is at capturing the strong-field effects we see.

Our results are summarized in Fig.~\ref{fig:mismatch}. An indicative number is $1-\mathfrak{M} \gtrsim 0.965$, corresponding to a loss of less than $10\%$ of the events that
could be potentially be detected by a ``perfect'' filter~\cite{Owen:1995tm,Berti:2007zu}; regions where this bound is violated are shaded in Fig.~\ref{fig:mismatch}. The non-relativistic model fails to describe b-EMRI waveforms in the strong field limit for $\theta\gtrsim \pi/8$. For smaller $\theta$, the mismatch is smaller for higher values of the radius $r_0$. This is because in the face-on limit the only difference between our waveforms and the non-relativistic model is the amplitude modulation due to helicity dependent scattering. Since this effect decreases as the SB is further from the BH, according to Eq.~\eqref{eq:helicity}, the corresponding decrease in the mismatch is not surprising. 

\noindent\textbf{\emph{Discussion.}} 
%
All the results we discussed and analyzed in this \emph{Letter} are borne out of first-principled calculations. This includes beaming, Doppler modulation, lensing, resonances with the SMBH, and helicity non-preserving scattering. All these features make the b-EMRI signal very rich and with ample potential of probing strong field gravity and the nature of compact objects in galactic nuclei. 

We have also established that the benchmark models for Doppler shifts in the literature are not suited for this type of systems for a wide range of observing angles. Thus, in order to extract the full potential out of future observations of b-EMRIs, further work must be devoted to understand their dynamics in the strong field. 

We note that a similar approach to describe b-EMRIs was recently proposed~\cite{Yin:2024nyz}. Although we recover the same frequencies~\eqref{eq:frequencies} reported in that work, we were unable to recover their results for waveforms or energy fluxes (which don't match known limits, see~\ref{sec:benchmarks}). This might be due to numerical errors or insufficient number of harmonics in their analysis. Likewise, the rich physics content that we just discussed is altogether absent from their study.

This work serves as a proof-of-concept of all the interesting phenomenology associated to b-EMRIs, providing a model to study this system from first principles. Our framework can be readily extended to incorporate more generic b-EMRI orbits and more relativistic secondary binaries.
Several avenues seem to be open for exploration, such as the effect of orbital resonances or the adiabatic evolution of the system through GW emission. A detailed study of detectability and parameter estimation of b-EMRIs in view of our results is also mandatory for future studies. 

\noindent\textbf{\emph{Acknowledgments.}} 
 We are indebted to João Luís Rosa for producing and providing the data for the orbiting hotspot, 
as well as to Giulia Cusin, and especially to Martin Pijnenburg, for explaining their findings on helicity-dependent lensing to us. 
 We further thank Enrico Barausse, Xian Chen, Marina De Amicis, Marta Cocco, Francisco Duque, Josh Mathews, Marta Orselli, Jan Steinhoff, Yucheng Yin, and Miguel Zumalacárregui for their comments on a draft of this manuscript. 
 We also acknowledge Diogo Ribeiro and Zhen Zhong for fruitful conversations. 
 We acknowledge the use of the Black Hole Perturbation Toolkit~\cite{BHPToolkit} for comparisons and benchmarking, as well as xAct~\cite{Martin-Garcia:2008ysv} for the analytical calculations. 
The Center of Gravity is a Center of Excellence funded by the Danish National Research Foundation under grant No. 184.
We acknowledge support by VILLUM Foundation (grant no.\ VIL37766) and the DNRF Chair program (grant no.\ DNRF162) by the Danish National Research Foundation.
VC acknowledges financial support provided under the European Union’s H2020 ERC Advanced Grant “Black holes: gravitational engines of discovery” grant agreement no.\ Gravitas–101052587.
JN was partially supported by FCT/Portugal through CAMGSD, IST-ID, projects
UIDB/04459/2020 and UIDP/04459/2020, and by the European Union’s Horizon 2020
research and innovation programme H2020-MSCA-2022-SE
through project EinsteinWaves, grant agreement no.\ 101131233.
MvdM acknowledges financial support provided under the European Union’s Horizon ERC Synergy Grant “Making Sense of the Unexpected in the Gravitational-Wave Sky” grant agreement no.\ GWSky–101167314.
Views and opinions expressed are however those of the authors only and do not necessarily reflect those of the European Union or the European Research Council. Neither the European Union nor the granting authority can be held responsible for them.
This project has received funding from the European Union's Horizon 2020 research and innovation programme under the Marie Sk{\l}odowska-Curie grant agreement No 101007855 and No 101131233, 
as well as from Funda\c{c}\~{a}o para a Ci\^{e}ncia e a Tecnologia under the project 2024.04456.CERN.
%
\bibliography{my_ref}
%
\renewcommand\thesection{S.\arabic{section}}
\renewcommand\theequation{S.\arabic{equation}} 
\renewcommand\thefigure{S.\arabic{figure}}     

\clearpage
\appendix*
\section*{SUPPLEMENTAL MATERIAL}

\renewcommand{\thesubsection}{{S.\arabic{subsection}}}
\setcounter{section}{0}

\setcounter{section}{0}
\setcounter{equation}{0}
\setcounter{figure}{0}
%
\subsection{Modeling the SB}
%
In this section, we go into more detail on how to model the SB in the Kerr spacetime of the SMBH using Dixon's formalism. 
%
\subsubsection{The Kerr solution} \label{sec:Kerr}
%
We consider a b-EMRI as depicted in Fig.~\ref{fig:b_EMRI} of the main text. Using Boyer-Lindquist coordinates $\{t,r,\theta,\phi \}$, the geometry of a SMBH with mass $M$ and angular momentum $Ma$ in vacuum is described by 
\beq
ds^2 &=& - \left(1-\frac{2Mr}{\Sigma}\right) dt^2 - \frac{4 M a r \sin^2 \theta}{\Sigma} dt d\phi+\frac{\Sigma}{\Delta} dr^2  \nonumber \\ 
& +& \Sigma d\theta^2 + (r^2 +a^2 
+ \frac{2 M r a^2 }{\Sigma} \sin^2 \theta) \sin^2 \theta \; d\phi^2 , \label{eq:kerr_metric}
\eeq
where
\be
\Sigma =r^2 + a^2 \cos^2 \theta \, , \quad \Delta = r^2 -2 M r + a^2
\ee
and $0\leq a  \leq M$. The inner and outer horizons are located at the roots of $\Delta$,
\begin{equation}
r_\pm = M \pm \sqrt{M^2 -a^2}\, .
\end{equation}

This spacetime admits circular timelike geodesics in the equatorial plane ($\theta=\pi/2$). Particles in such orbits have 4-velocity 
\beq 
u^\mu= u^t \left(\delta^\mu _t + \Omega_{0}
\, \delta^\mu _\phi \right) \, , \label{eq:geodesic}
\eeq
with orbital frequency 
\beq 
\Omega_0 = \frac{\pm\sqrt{M}}{r_0 ^{3/2} \pm a \sqrt{M}} \, , \label{eq:omega_geo}
\eeq
where the plus and minus sign refers to prograde and retrograde orbits, respectively. The quantity $u^t$ is determined by the normalization condition $u^\mu u_\mu = -1$.

\subsubsection{Dixon's formalism} \label{sec:Dixon}
%
A generic matter distribution, free falling in curved spacetime and described by an energy momentum tensor $T^{\mu \nu}$ has the simple equation of motion
$$ \nabla_\mu T^{\mu \nu} = 0 \, .$$
The Dixon formalism~\cite{Dixon:1970zza,Dixon:1974xoz,Dixon:2015vxa} provides a simpler description of the problem for systems where the matter distribution under consideration has a typical length scale smaller than the local radius of curvature of the ambient spacetime, $d \ll \sqrt{r^3/M}$. In this approach, an extended mass distribution (in our case the SB) is described via its gravitational skeleton, i.e.\ its multipolar structure. We will keep terms up to quadrupolar order, that is, $\mathcal{O} (d /  \sqrt{r^3/M})^2$.

The mass distribution is taken to be supported in a worldtube, constructed around a reference worldline $z^\mu(\tau)$, where $\tau$ is the proper time along the curve. One then considers a foliation of the worldtube by spacelike surfaces $\Sigma_\tau$ which are generated by the geodesics emanating from $z^\mu(\tau)$ and orthogonal to $u^\mu \equiv dz^\mu / d\tau$ at that point~\cite{Dixon:2015vxa,Harte:2011ku,Costa:2012cy, Gralla:2010xg}. The moments of the mass distribution are obtained by taking integrals over the leaves of the foliation using the theory of bitensors~\cite{Poisson:2011nh,Harte:2011ku}. 

Dixon's formalism yields evolution laws for the momentum $p_\mu (\tau)$ (monopole) and spin tensor $S^{\mu \nu} (\tau) = S^{[\mu \nu]} (\tau)$ (dipole). At quadrupolar order, for a free-falling mass distribution, these take the form 
\begin{align}
    \frac{D p_\mu}{d \tau} & =  -\frac{1}{2}R_{\mu \nu \rho \sigma}u^\nu S^{\rho \sigma} - \frac{1}{6} J^{\nu \rho \sigma \delta} \nabla_\mu R_{\nu \rho \sigma \delta}  \, , \label{eq:MPD_1}\\ 
    \frac{D S^{\mu \nu}}{d \tau} & = 2 p^{[\mu} u ^{\nu]}- \frac{4}{3} J ^{ \rho \sigma \delta [\mu} R^{\nu]}_{\ \ \rho \sigma \delta} \, , \label{eq:MPD_2}
\end{align}
where $u^a (\tau)=d z^a/ d\tau$ is the tangent vector to the reference worldline and $J^{\alpha\beta\gamma\delta} (\tau)$ is the reduced quadrupole moment. The quadrupole moment shares the symmetries of the Riemann tensor, namely 
\beq
     & & J^{\mu \nu\rho \sigma} = - J^{\nu \mu \rho \sigma} = - J^{\mu \nu \sigma \rho} \nonumber \, , \\
    & & J^{\rho \sigma\mu \nu} = J^{\mu \nu\rho \sigma} \, , \label{eq:sym} \\ 
    & & J^{[\mu \nu \rho]\sigma} = 0 \nonumber \, .
\eeq

The system~\eqref{eq:MPD_1}-\eqref{eq:MPD_2} is undetermined, and must be closed by choosing a spin supplementary condition. This choice is equivalent to choosing the observer who is measuring the center of mass to be on the reference worldline~\cite{Costa:2014nta}. In our case, the spin supplementary condition is very simple and is given in Eq.~\eqref{eq:SSC}. 

It is worth noting that $p^\mu$ is not necessarily proportional to $u^\mu$. The exact relation between the two depends on the choice of supplementary condition; however, in our modeling of the system we do have $p^\mu\propto u^\mu$. Finally, this formalism does not give evolution equations for the multipoles beyond the spin. Namely, the quadrupole moment must be calculated independently. 

These equations allow us to study the dynamics of the extended body as if it were a point particle located at $z(\tau)$, but with internal structure given by its multipole moments. A natural follow-up question is: can we perform a similar \textit{skeletonization} of the energy-momentum tensor? In other words, can we obtain a $T^{\mu \nu} (\tau)$ with support on the worldline that is, to the given multipolar order, equivalent to the original energy-momentum tensor? The answer is yes, and the result is Eq.~\eqref{eq:canonical_EM_tensor}.
%
\subsubsection{Dynamics of the secondary binary} \label{sec:SB_dynamics}
%
As stated in the main text, we take the center of mass of the SB to move along a circular geodesic. To understand why, consider an orbit of the SB around the SMBH with radius $r_0$ and orbital frequency $\Omega_0$. We are interested in the GW signal from the triple system over timescales comparable to the orbital period $2\pi /\Omega_0$. On that scale, we can consider as a first approximation 
\begin{equation}
    \frac{D p^\mu}{d \tau}= \frac{D u^\mu}{d \tau} = 0 \, , \quad p^\mu = 2 \mu u^\mu \, . \label{eq:MPD_geo_1}
\end{equation}
In this approximation, we are neglecting the terms on the RHS of Eq.~\eqref{eq:MPD_1} and the radiation reaction force~\cite{Poisson:2011nh}, $F_{\rm RR}$, which are at most
\begin{align}
    &R_{\mu \nu \rho \sigma}u^\nu S^{\rho \sigma} \sim \frac{M\mu d}{r_0 ^3}  \nonumber  \, ,\\
    &J^{\nu \rho \sigma \delta} \nabla_\mu R_{\nu \rho \sigma \delta} \sim \frac{\mu M d^2 }{r_0^4}  \, ,\\ 
    &F_{\rm RR} \sim \mu^2 r_0 ^4 \Omega_0^6 \nonumber \, ,
\end{align}
all of which are $\ll \mu \, \Omega_0$. This approximation also yields that the spin tensor is parallel-transported,
\begin{equation}
    \frac{D S^{\mu \nu}}{d \tau}  = 0 \, . \label{eq:MPD_geo_2}
\end{equation}
Since $p^\mu$ is parallel to $u^\mu$, we can simultaneously implement the Tulczyjew-Dixon and Mathisson-Pirani spin supplementary conditions~\cite{Costa:2014nta}, that is, 
\begin{equation}
    S^{\mu \nu}u_\nu = S^{\mu \nu}p_\nu = 0 \,. \label{eq:SSC}
\end{equation}

Given this condition, it is natural to define a spin vector orthogonal to the 4-velocity, given by
\begin{equation}
    S^\mu = \frac{1}{2} \epsilon^{\mu \nu \rho \sigma} u_\nu S_{\rho \sigma} \, , \label{eq:spin_vector}
\end{equation}
where $\epsilon^{\mu \nu \rho \sigma}$ is the Levi-Civita tensor with $\epsilon^{0123}=\sqrt{-g}$. 

From here on, we take the SB to be moving in a circular geodesic of radius $r_0$ in the equatorial plane of the SMBH, so $u^\mu$ can be read off from Eqs.~\eqref{eq:geodesic} and~\eqref{eq:omega_geo}. We allow for generic inclinations of the spin of the SB relative to the angular momentum of the SMBH. Furthermore, we will model the interior dynamics of the SB using purely Newtonian mechanics, limiting our analysis to the equal-mass circular orbit case. Thus, we need only calculate the moments of the SB for the configuration described above. This is more easily done by transforming to an appropriate frame. 
\begin{figure*}[ht!]
\centering
\begin{subfigure}{0.3 \textwidth}
\centering
\includegraphics[width=.66\textwidth]{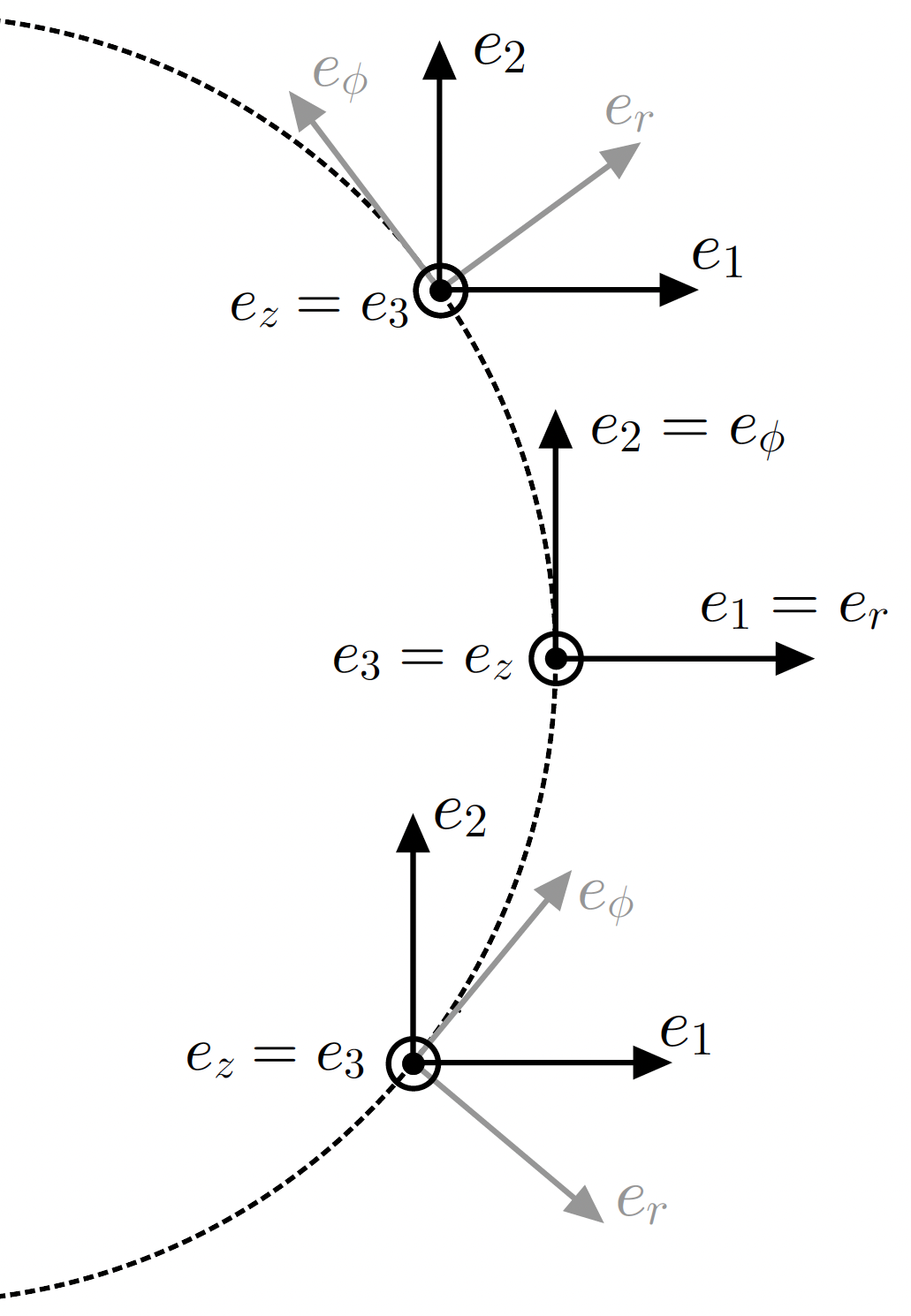}
\end{subfigure}
\hspace{0.01 \textwidth}
\begin{subfigure}{0.3\textwidth}
\centering
\includegraphics[width=\textwidth]{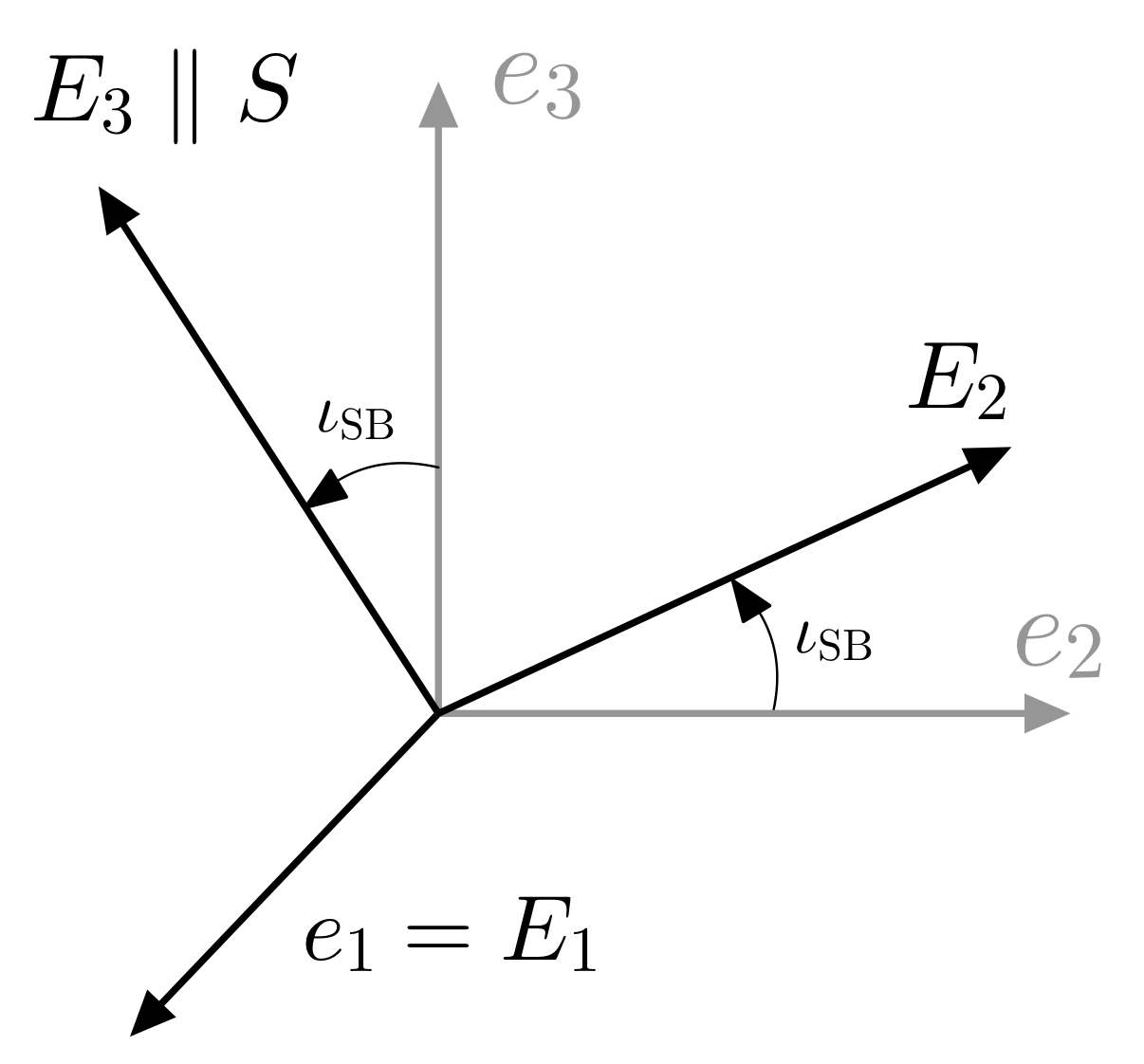}
\end{subfigure}
\hspace{0.01 \textwidth}
\begin{subfigure}{0.3\textwidth}
\centering
\includegraphics[width=.9 \textwidth]{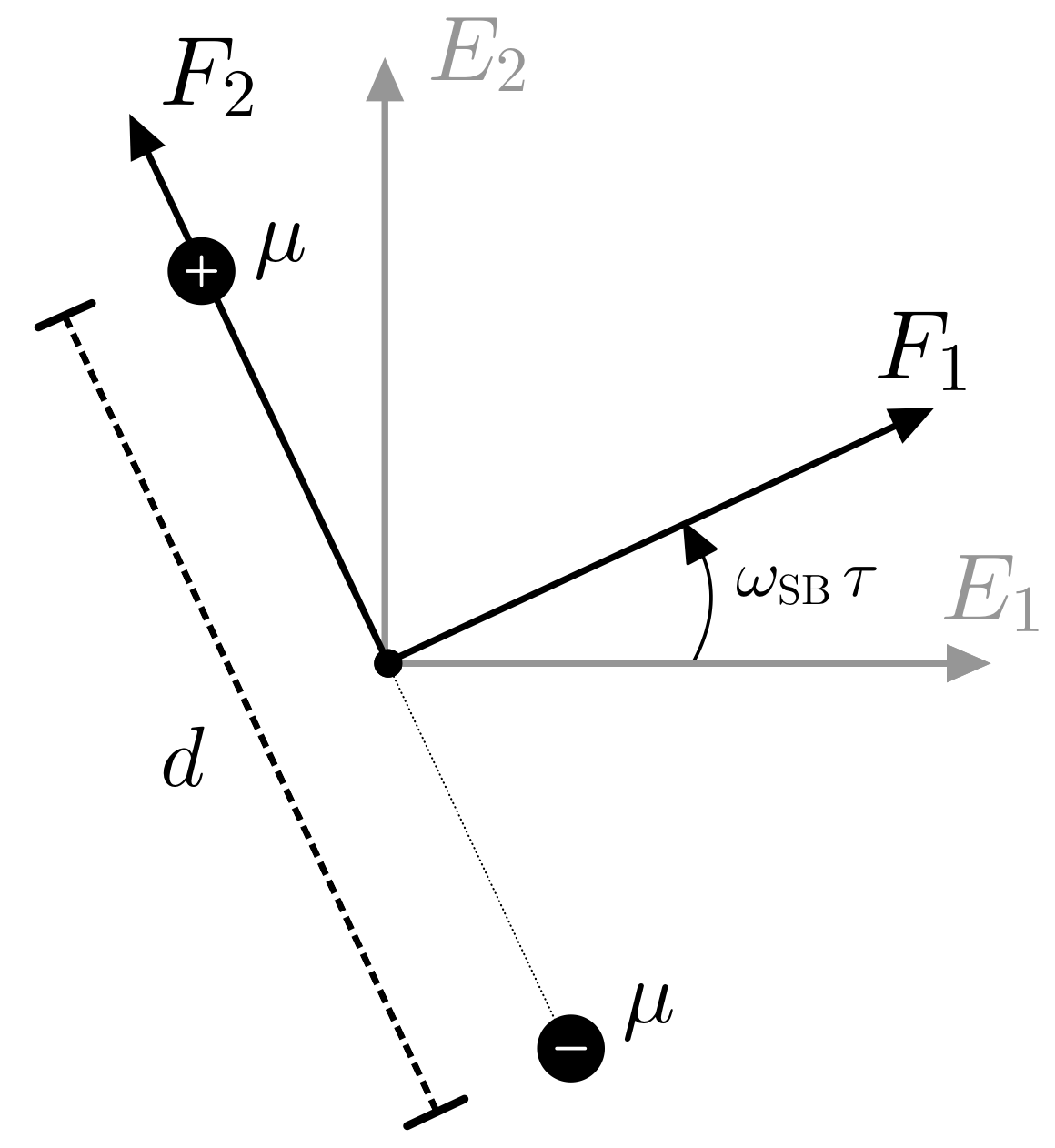}
\end{subfigure}
\caption{\justifying Schematic representation of the transformations between the frames used to calculate the moments of the SB. \emph{Left:} Local inertial frame (gray -- see Eqs.~\eqref{eq:LIF1}--\eqref{eq:LIF3})$\to$ parallel-transported frame (black -- see Eqs.~\eqref{eq:PTF1}--\eqref{eq:PTF3}); the dashed line denotes the circular geodesic in the equatorial plane followed by the SB. \emph{Center:} Parallel-transported frame (gray) $\to$ spin-aligned frame (black -- see Eqs.~\eqref{eq:SAF1}--\eqref{eq:SAF3}); the Euler angle used to align $E_3$ with the spin of the SB, $S$, is the inclination $\iota_{\rm SB}$. \emph{Right:} Spin-aligned frame (gray) $\to$ binary-aligned frame (black -- see Eqs.~\eqref{eq:BAF1}--\eqref{eq:BAF2}); the two components of the SB (black dots with mass $\mu$ and separated by a distance $d$) are aligned with the $F_2$ axis (and identified with $\pm$ signs depending on whether they are in the $\pm F_2$ direction). The orbital frequency of the SB, with respect to proper time $\tau$ along the circular geodesic, is ${\rm\omega_{SB}}$.}\label{fig:frames}
\end{figure*}
%
\subsubsection{Frame transformations} \label{sec:frames}
%
The moments of the SB can be most easily calculated by exploring the fact that spacetime is locally described by the Minkowski metric. We start by introducing a local inertial frame attached to the SB with basis vectors $\{u, e_r,e_\phi, e_z \}$, whose components are
\beq
    & & u^\mu = u^t(\delta^\mu _t + \Omega_0 \delta^\mu  _\phi) \label{eq:LIF1} \, , \\ 
    & & e_r ^\mu = \sqrt{\frac{\Delta}{\Sigma}} \delta^\mu _r \, , \\
    & & e_z ^\mu = - \sqrt{\frac{1}{\Sigma}}\delta^\mu_\theta \, , \label{eq:LIF3}
\eeq
where we should remember that all these quantities are to be evaluated at the SB orbit, that is for $r=r_0$, $\theta = \pi/2$. The fourth vector in the tetrad, $e_\phi$, is defined by requiring orthonormality. 

Next, consider a frame which is parallel-transported along the geodesic (see left panel in Fig~\ref{fig:frames}). When a vector $e$ orthogonal to the 4-velocity is parallel-transported along a circular equatorial geodesic, it precesses with respect to the triad $\{e_r, e_\phi, e_z\}$ according with the equation \cite{Rindler_Perlick_1990, vandeMeent:2019cam}
\beq
    \frac{d e}{d \tau} = {\rm \omega_P} \ e\cross e_z =  \pm \sqrt{\frac{M}{r_0^3}} \ e \cross e_z \, ,
\eeq
where ${\rm \omega_P}$ is the precession frequency (in proper time) and the top and bottom signs refer to prograde and retrograde orbits, respectively. Thus, the parallel-transported frame $\{u, e_1, e_2, e_3 \}$ is simply
\begin{align}
    & e_1 = \cos \left(-{\rm \omega_P} \tau\right) \ e_r + \sin \left(-{\rm \omega_P} \tau\right) \ e_\phi \, , \label{eq:PTF1}\\
    & e_2  = -\sin \left(-{\rm \omega_P} \tau\right) \ e_r + \cos \left(-{\rm \omega_P} \tau\right) \ e_\phi \, , \\
    & e_3 = e_z  \, . \label{eq:PTF3}
\end{align}

Now we want to introduce a frame which is also parallel-transported, but where one of the vectors of the tetrad, say $E_3$, is aligned with the spin vector of the SB (see center panel in Fig~\ref{fig:frames}). Note that since the spin vector is parallel-transported, if this condition is met initially then it will always be met. As the spin vector precesses at a constant frequency ${\rm \omega_P}$, we only have to specify the inclination of $S$ relative to $e_3$. We can do this with a single Euler angle, the inclination $\iota_{\rm SB}$. The resulting frame $\{u, E_1, E_2, E_3 \}$ is then
\begin{align}
   & E_1  =  e_1  \, , \label{eq:SAF1}\\
   & E_2  = \cos{\iota_{\rm SB}} \ e_2 + \sin{\iota_{\rm SB}}\  e_3 \, , \\
   & E_3  = - \sin{\iota_{\rm SB}}\ e_2 + \cos{\iota_{\rm SB}} \ e_3  \, . \label{eq:SAF3}
\end{align}
By construction, in this frame the SB is moving counterclockwise in the plane spanned by $\{E_1,E_2 \}$. 

We now focus on the inner dynamics of the SB. For this, we use Newtonian physics and consider the equal mass circular orbit case. If the two components of the SB have the same mass $\mu$ and are separated by a distance $d$, then the orbital frequency is 
\begin{equation}
    {\rm \omega_{SB}} = \sqrt{\frac{2 \mu}{d^3}} \, . \label{eq:OmegaSB}
\end{equation}
Naturally, this is the orbital frequency in proper time $\tau$. 

Now consider a final frame $\{ F_1, F_2\}$ on the $\{ E_1, E_2\}$ plane such that the two components of the binary are always on the $F_2$ axis (see right panel in Fig~\ref{fig:frames}). This frame is given by 
\begin{align}
    F_1  =& \cos{\left({\rm \omega_{SB}} \tau \right)} \ E_1 + \sin{\left({\rm \omega_{SB}} \tau \right)} \ E_2      \, , \label{eq:BAF1}\\
    F_2  =& -\sin{\left({\rm \omega_{SB}} \tau \right)} \ E_1 + \cos{\left({\rm \omega_{SB}} \tau \right)} \ E_2      \, . \label{eq:BAF2}
\end{align}
%
%
\subsubsection{Quadrupole moment of an equal mass binary in flat space} \label{sec:flat}
%
Take a fixed value of $\tau$ corresponding to a point on the orbit $z^\mu(\tau)$. Consider a small neighborhood of $z^\mu(\tau)$ containing the SB but still of size $\sim d \ll \sqrt{r_0^3 /M}$. We can parameterize this neighborhood using Riemann normal coordinates~\cite{Poisson:2011nh} associated to the frame $\{u, F_1, F_2, E_3 \}$, denoted by $\{\hat x^0, \hat x^1,\hat x^2, \hat x^3 \}$. Given the scales mentioned above, this region is well described as a neighborhood of the origin in Minkowski space with coordinates $\{\hat x^0, \hat x^1,\hat x^2, \hat x^3 \}$. 

The multipole moments are defined on $z^\mu(\tau)$, and are obtained from integrals over spacelike geodesic surfaces orthogonal to the worldline $z^\mu(\tau)$ at each point. In this simplified picture, this means the hyperplane $\hat x_0 = 0$, which is represented in the rightmost image in Fig.~\ref{fig:frames}. As indicated in the figure, we use a ``$+$" (resp.\ ``$-$") to refer to the component of the SB with $\hat{x}^2>0$ (resp. $\hat{x}^2<0$).

We make the further approximation that the SB components are point particles of mass $\mu$, so that their 4-velocities are just
\begin{equation}
    U^{\hat\mu} _\pm = \delta^{\hat\mu} _0 \mp \left( d \, {\rm \omega_{SB}} / 2  \right) \,  \delta^{\hat\mu} _1 \, , \quad d \, {\rm \omega_{SB}} / 2  \ll 1 \, , \label{eq:U_component}
\end{equation}
where we introduce hatted indices to indicate expressions in the frame $\{u, F_1, F_2, E_3 \}$. From the 4-velocities one easily obtains the energy-momentum tensor for each of the particles,
\begin{equation}
    T^{\hat\mu \hat\nu} _\pm = \mu U^{\hat\mu} _\pm U^{\hat\nu}_\pm \delta(\hat{x}^1)\delta(\hat{x}^2\mp d/2)\delta(\hat{x}^3) \, . \label{eq:EMT_component}
\end{equation}
The full energy-momentum tensor of the SB is then $T^{\hat\mu \hat\nu} _{\rm SB} = T^{\hat\mu \hat\nu} _{+} \, + T^{\hat\mu \hat\nu} _{-}$. We can now calculate the moments of the SB, which take a simple form in our coordinates~\cite{Harte:2020ols}:
\begin{align}
    p^{\hat{\mu}} =& \int_{\mathbb{R}^3} T_{\rm SB}^{\hat{\mu} 0} d^3 \hat{x}\, , \nonumber \\ 
    S^{\hat\mu \hat\nu} =&  2 \int_{\mathbb{R}^3} \hat{x}^{[\hat{\mu}} T_{\rm SB}^{\hat{\nu}] 0} d^3 \hat{x}\, , \label{eq:multipoles} \\ 
    J^{\hat\mu \hat\nu \hat\rho \hat\sigma} =& \int_{\mathbb{R}^3} \hat{x}^{[\hat\mu}\big(T_{\rm SB}^{\hat\nu][\hat\sigma} + T_{\rm SB}^{\hat\nu]0}\delta^{[\hat\sigma}_0 + \delta^{\hat\nu]}_0 T_{\rm SB}^{0[\hat\sigma}\big)\hat x ^{\hat\rho]} \ d^3 \hat{x} \, , \nonumber
\end{align}
for the momentum (monopole), spin (dipole) and quadrupole tensors, respectively. The integration over $\mathbb{R}^3$ is an abuse of notation, but it is not important since the energy-momentum tensor has support in a region where spacetime is indeed approximately flat. Plugging the explicit form of $T^{\hat\mu \hat\nu} _{\rm SB}$ in these definitions we obtain
\begin{align}
    & p^{\hat{\mu}} = 2 \mu \, \delta_0 ^{\hat\mu}\, , \nonumber \\ 
    & S^{\hat\mu \hat\nu} =   \mu d^2 {\rm \omega_{SB}} \, \delta_1 ^{[\hat \mu} \delta_2 ^{\hat \nu]}\, , \label{eq:multipoles_explicit} \\ 
    & J^{\hat\mu \hat\nu \hat\rho \hat\sigma} = \frac{3}{8} \mu d^2 \, \delta_0^{\hat\mu} \delta_2^{\hat\nu} \delta_0^{\hat\rho} \delta_2^{\hat\sigma}+\frac{1}{32} \mu d^4 {\rm\omega_{SB}}^2\, \delta_1^{\hat\mu} \delta_2^{\hat\nu} \delta_1^{\hat\rho} \delta_2^{\hat\sigma}+\text{(sym)}  \, , \nonumber
\end{align}
where (sym) indicates all the other terms obtained from applying the symmetries of the quadrupole tensor in Eq.~\eqref{eq:sym} to the first two terms. Note that indeed we obtained $p^\mu = 2\mu u^\mu$, where $u_\mu$ is the 4-velocity of the SB. Moreover, the spin vector (see Eq.~\eqref{eq:spin_vector}) is simply 
\begin{equation}
    S^{\hat\mu} = \frac{1}{2} \mu  d \, {\rm\omega_{SB}} \, \delta^{\hat\mu}_3 \, ,
\end{equation}
which coincides with the Newtonian angular momentum of the system.

To obtain the components of the multipole moments in the Boyer-Lindquist frame, we simply have to perform the frame transformations defined in Sec.~\ref{sec:frames}. Since some of these transformations depend on $\tau$, the multipole moments will be functions of $\tau$.

Note that the geodesic approximation taken in Eqs.~\eqref{eq:MPD_geo_1} and~\eqref{eq:MPD_geo_2} is consistent  with calculating the GWs produced by this system up to quadrupole order. However, for timescales longer than the orbital period $2 \pi / \Omega_0$, where self-force effects are important, the force terms in the RHS of Eqs.~\eqref{eq:MPD_1} and~\eqref{eq:MPD_2} must be accounted for. Such extension is a straightforward generalization of this work, and amounts to accounting for the adiabatic evolution of Dixon's moments through GW emission. 
%
%
%
\subsection{Gravitational wave generation}
%
%
In the previous section, we established the multipole moments up to quadrupole order of the SB in the SMBH spacetime. We now describe the formalism to compute GWs from the b-EMRI system. Since the mass of the SB components satisfies $\mu\ll M$, where $M$ is the mass of the SMBH, we are fully in the realm of perturbation theory, so we will use the Teukolsky formalism.
%
%
\subsubsection{The Teukolsky equation} \label{sec:teukolsky}
%
The Teukolsky master variable and matter fields introduced in the main text take the form 
\begin{align}
    \psi \equiv & \rho^{-4} \psi_4 \label{eq:fourier_harmonic}  = \int _{\mathbb{R}} \ \sum _{\ell,m } ^\infty R_{\ell m \omega } (r) _{-2}
 S_{\ell m \omega} (\theta) e^{i m \phi} e^{-i \omega t}  d \omega \, , \\
    \mathcal{T} \equiv & 2  \rho^{-4} T_4 =  \int _{\mathbb{R}} \ \sum _{\ell,m } ^\infty \tilde T_{\ell m \omega } (r)  _{-2}
 S_{\ell m \omega } (\theta) e^{i m \phi} e^{-i \omega t}  d \omega \, , \nonumber 
\end{align}
where $\rho = -(r-i a \cos\theta )^{-1}$,  $\psi_4$ is a particular projection~\cite{Teukolsky:1973ha} of the Weyl tensor onto the Kinnersley tetrad (cf.~Eq.~\eqref{eq:Kinnersley}), and the form of $T_4$ will be discussed below. 
Separation of variables yields two ordinary differential equations per mode, one in the radial coordinate $r$ and another in the $\theta$ coordinate. The angular equation is a regular Sturm-Liouville problem with eigenvalues $_s E_{\ell m \omega }$, and defines spin-weighted spheroidal harmonics.  There is no closed form analytical expression for the eigenvalues or eigenfunctions of the angular equation: they must be computed numerically~\cite{Press:1973zz,1986JMP....27.1238L,Berti:2005gp}.

The radial equation is a Schrödinger-like equation; it takes the form, omitting the subscripts in $R$ and $T$ for simplicity,
\beq
&&\Delta^{2} \partial_r \lr{\Delta^{-1}\partial_r R} - V(r) R = - 4 \pi \Sigma \tilde T\,,\label{eq:teukolsky_radial}\\
&&V(r) = \frac{K^2 +4i (r-M) K}{\Delta} -8 i  \omega r - \lambda \, , \nonumber
\eeq
where $K= (r^2 + a^2)\omega - a m  $ and $\lambda= {_{-2} E_{\ell m \omega}} - 2 a m \omega + a^2 \omega^2 -2 $. We need to solve the radial equation~\eqref{eq:teukolsky_radial} with appropriate boundary conditions, corresponding to purely ingoing waves at the horizon and purely outgoing waves at infinity. This is done by first constructing two linearly independent solutions -- $R^H$ and $R^\infty$ -- to the homogeneous problem ($T=0$), with the following asymptotics~\cite{Teukolsky:1973ha}:
\beq
R^H &\sim & A_{\text{in}}\  r^{-1}\ e^{- i \omega r_\star} + A_{\text{out}} \ r^{3} \ e^{ i \omega r_\star} \, , \nonumber \\
R^\infty &\sim&  r^{3} \ e^{i \omega r_\star}\qquad (r\to\infty)\,, \label{eq:asymptotics_inf} \\ \vspace{0.5 cm}
R^\infty &\sim & B_{\text{in}} \ \Delta^{2} e^{- i k r_\star} + B_{\text{out}}\ e^{ i k r_\star} \, ,  \nonumber \\
R^H & \sim & \Delta^{2} \ e^{- i k r_\star}\qquad (r\to r_+)\,, \label{eq:asymptotics_hor}
\eeq
where $r_\star$ is the usual tortoise coordinate in the Kerr metric and $k = \omega - m \Omega_H$, for $\Omega_H = a/2 M r_+$ the angular velocity of the Kerr BH's horizon~\cite{Teukolsky:1973ha}. From these solutions we can then construct a rescaled Wronskian (which is independent of $r$),
\begin{align}
    W&=\Delta^{-1} \lr{R^\infty \frac{d R^H 
}{d r}-R^H \frac{d R^\infty}{d r}}\\
&= -2 i \omega A_{\rm in}
\,. \label{eq:Wronskian}
\end{align}
In the asymptotic regions at infinity and the BH horizon, the solution to the non-homogeneous problem ($T\neq0$) is then simply
\begin{equation}
    R \sim \left\{
    \begin{aligned}
        &Z_{\ell m\omega}^{\infty} r^{3} \ e^{i \omega r_\star}\qquad &&(r\to\infty) \\
        &Z_{\ell m\omega}^{H}\Delta^{2} \ e^{- i k r_\star}\qquad &&(r\to r_+)
    \end{aligned} 
    \right. \quad  ,
    \label{eq:gen_sol_r}
\end{equation}
with the amplitudes
\begin{align}
    Z_{\ell m\omega}^{\infty,H} &=  \frac{1}{W}  \int_{r_+} ^\infty \frac{4 \pi \Sigma}{\Delta^{2}} R ^{H,\infty} (r^\prime) \tilde T(r^\prime) \ d r^\prime \, .
    \label{eq:amplitudes_SM}
\end{align}

\subsubsection{The source term $T_4$} \label{sec:source}
%
The source term $T_4$ is constructed from the energy-momentum tensor of the matter in the vicinity of the SMBH, in our case~\eqref{eq:canonical_EM_tensor}. To write it explicitly, we first introduce the Kinnersley tetrad, $\{l, n, m, \bar m \}$, where we have~\cite{Kinnersley:1969zza, Teukolsky:1973ha}  
\begin{align}
		&l^\mu =\left(\frac{r^2 + a^2}{\Delta},1,0,\frac{a}{\Delta} \right), \nonumber \\
		&n^\mu =\frac{1}{2\Sigma}\left(r^2 +a^2,-\Delta,0,a  \right) \, , \label{eq:Kinnersley} \\
		&m^\mu =\frac{1}{\sqrt{2}\left(r + i a \cos\theta\right)}\left(i a \sin\theta,0,1,i/ \sin \theta  \right) , \nonumber
\end{align}
and a bar denotes complex conjugation. The source term $T_4$ is simply 
\begin{equation}
    T_4 \equiv \mD_{nn} T_{nn} + \mD_{n \bar m} T_{n \bar m} + \mD_{\bar m \bar m} T_{\bar m \bar m} \, ,  \label{eq:T4}
\end{equation}
where $\mD_{\bullet \bullet}$ denote second-order differential operators obtained from terms like $n^\mu \partial_\mu$ and spin coefficients (for explicit expressions see Eq.(2.15) of Ref.~\cite{Teukolsky:1973ha}) and $T_{\bullet \bullet}$ are contractions of $T_{\mu \nu}$ with the Kinnersley tetrad. To translate this to the Fourier-harmonic amplitudes of Eq.~\eqref{eq:fourier_harmonic}, we use the orthogonality of the spin weighted spheroidal harmonics and write 
\begin{equation}
    \tilde T_{\ell m \omega} = \frac{1}{\pi} \int_{\mathbb{R}} e^{i \omega t^\prime } \int_{\mathbb{S}^2} \, _{-2} \bar S_{\ell m \omega} \, e^{- i m \phi^\prime} \, \frac{ T_4}{\rho^{4}} \   d\Omega^\prime d t^\prime \, , \label{eq:T_amplitude}
\end{equation}
where $d\Omega^\prime = \sin \theta^\prime d \theta^\prime d \phi^\prime $ is the standard volume element on $\mathbb{S}^2$. Replacing this result in Eq.~\eqref{eq:amplitudes_SM}  yields, omitting the subscripts
\begin{align}
  & Z^{\infty,H} = \int_{\mathcal{M}} {\frak F}^{\infty,H} \, e^{i(\omega t^\prime - m\phi^\prime)} T_4(x^\prime) \ d^4 x^\prime \quad\, , \label{eq:amplitude_inf} \\ 
   &{\frak F}^{\infty,H}(r^\prime, \theta^\prime) = \frac{4\, \Sigma \sin \theta^\prime}{W \Delta^2 \rho^4} R^{H,\infty} (r^\prime) \,   \bar S(\theta^\prime) \, , \nonumber
\end{align}
where $\mathcal{M} = \mathbb{R} \times (r_+ , \infty)\times\mathbb{S}^2 $ and the volume element is $d^4 x^\prime =d\phi^\prime d\theta^\prime d r^\prime d t^\prime $.
%
\subsubsection{The amplitude for a b-EMRI} \label{sec:source_B_EMRI}
%
Let us now focus on the actual source that we are studying -- the SB with $T^{\mu \nu}$ given by~\eqref{eq:canonical_EM_tensor}. The method to obtain the multipole moments was discussed in the last section of the SM. The energy-momentum tensor is supported on the worldline of the SB, which we take to be a circular equatorial geodesic of the Kerr geometry, that is, in Boyer-Lindquist coordinates
\begin{equation}
    z(\tau) = (u^t \tau\, ,\,  r_0\, ,\,  \pi/2 \, , \, u^t 
 \Omega_0 \tau ) \, , 
\end{equation}
where $\Omega_0$ is given by a choice of sign in Eq.~\eqref{eq:omega_geo}, depending on wether the orbit is prograde or retrograde. 

For clarity, we will build the amplitude in Eq.~\eqref{eq:amplitude_inf}, which depends on $T^{\mu\nu}$, from pieces which depend only on certain terms of the energy-momentum tensor, namely 
\begin{align}
   {\overset{0}{T}} \hspace{0 em} ^{\mu \nu} (x) =& \int d\tau u^{(\mu} p ^{\nu)} \delta_{(4)} \label{eq:canonical_EM_tensor_0} \, , \\
    {\overset{1}{T}} \hspace{0 em} ^{\mu \nu} (x) =& - \int d \tau \ \nabla_\rho \left(S^{\rho (\mu} u^{\nu)} \delta_{(4)} \right) \label{eq:canonical_EM_tensor_1} \, , \\
    {\overset{2}{T}} \hspace{0 em} ^{\mu \nu} (x) =& - \frac{2}{3} \int d \tau \ \nabla_\sigma \nabla_\rho \left(J^{\sigma(\mu \nu)\rho} \delta_{(4)} \right) \, ,\label{eq:canonical_EM_tensor_2} \\
    {\overset{3}{T}} \hspace{0 em} ^{\mu \nu} (x) =& \frac{1}{3} \int d\tau   R_{\rho \sigma \delta}^{\quad (\mu}J^{\nu) \delta \sigma \rho} \delta_{(4)} \label{eq:canonical_EM_tensor_3} \, .
\end{align}
We label each of these terms by $k\in \{0,1,2,3\} $, and call them, respectively, the monopole, dipole, dynamic quadrupole, and tidal quadrupole. In~\ref{app:Newtonian_quadrupole} we show that by taking the Newtonian limit we recover the standard energy momentum tensor at quadrupole order of a Newtonian equal mass binary~\cite{Peters:1963ux,Poisson_Will_2014}. The contributions from the first and last terms are the easiest to calculate as they involve no covariant derivatives of the delta function. The other two are progressively more involved, as we will see below. 
%

%
\noindent{\emph{$\overset{0}{T} \hspace{0 em} ^{\mu \nu}, \overset{3}{T} \hspace{0 em} ^{\mu \nu} $: zero covariant derivatives.}}\label{sec:source_0}
%
In Eq.~\eqref{eq:T4} there are differential operators acting on the energy-momentum tensor. In order to integrate over the spacetime in Eq.~\eqref{eq:amplitude_inf}, we must remove the derivatives acting on $\delta_{(4)}$. This problem is common to all values of $k$, and is solved by integrating by parts, yielding
\begin{align}
    \kZ^{\infty,H} = \int_{\mathcal{M}} \bigg( &\overset{k}{T} \hspace{0 em}_{nn} \mD_{nn} ^\ast + \overset{k}{T} \hspace{0 em}_{n \bar m}\mD_{n \bar m} ^\ast +  \overset{k}{T} \hspace{0 em}_{\bar m \bar m} \mD_{\bar m \bar m} ^\ast \bigg)\label{eq:amplitude_inf_0} \\ 
    &\times {\frak F}^{\infty,H} \, e^{i(\omega t^\prime - m\phi^\prime) } \ d^4 x^\prime  \quad k\in\{0,1,2,3\} \, . \nonumber
\end{align}
Here the operators $\mD_{\bullet \bullet}^\ast$ are the formal adjoints of the operators $\mD_{\bullet \bullet}$. In~\ref{app:formal_adjoints} we explicitly obtain $\mD^\ast _{nn}$, and the calculation easily generalizes to the other terms. 

Focusing again on $k\in\{0,3 \}$, the delta functions have no derivatives acting on them, and so the integration can be carried out. However, contrary to the point particle case~\cite{Detweiler:1978ge}, now $T^{\mu \nu}$ contains explicit time dependencies. These appear when changing from the local frame, where Eq.~\eqref{eq:multipoles_explicit} was obtained, to the Boyer-Lindquist frame. Thus, we write all time dependencies as complex exponentials. For example, Eq.~\eqref{eq:BAF1} is rewritten as
\begin{equation*}
    F_1  = \frac{e^{i {\rm \Omega_{SB}} t}+ e^{- i {\rm \Omega_{SB}} t }}{2} \ E_1 + \frac{e^{i {\rm \Omega_{SB}}t}- e^{- i {\rm \Omega_{SB}} t }}{2 i}  \ E_2      \, , \\
\end{equation*}
where $\Omega_{\rm SB} = \omega_{\rm SB}/ u^t$ is the intrinsic frequency of the SB with respect to Boyer-Lindquist time, introduced in Eq.~\eqref{eq:frequencies_BL}. We recast all other time dependent frame transformations in a similar way, also introducing $\Omega_{\rm P}=\omega_{\rm P}/u^t$. The integrand in Eq.~\eqref{eq:amplitude_inf} becomes a sum of terms, each having the time dependence only in the form of a phase. There are three frequencies in the problem: the orbital frequency $\Omega_0$, the precession frequency ${\rm \Omega_P}$, and the frequency of the SB inner motion ${\rm \Omega_{SB}}$. Thus, given the form of the frame transformations, these phases will be of the form $\exp\left( - i (\omega-\omega_{mpq})t\right)$, for $\omega_{mpq}$ as given in Eq.~\eqref{eq:frequencies}. The system then has a total of fifteen frequencies for every value of $m$, all of which can be labeled by triplets $(m,p,q)$. The monopole term~\eqref{eq:canonical_EM_tensor_0} excites only the $(m,0,0)$ frequency, while the tidal quadrupole term~\eqref{eq:canonical_EM_tensor_3} excites all the frequencies in the problem. In general, the amplitude is 
\begin{equation}
    \kZ^{\infty,H}_{\ell m \omega } = \sum_{\substack{p\in\{0,\pm1,\pm2\} \\ q\in\{0,\pm2\}}} \kZ^{\infty,H}_{\ell m p q} \delta_{(mpq)} \,,
    \label{eq:amplitude_inf_0_integrated}
\end{equation}
for $k \in \{0,1,2,3\}$, and where $\delta_{(mpq)} = \delta(\omega - \omega_{m p q})$. The amplitudes $\kZ^\infty_{\ell m p q}$ only depend on the parameters of the system, and take the general form
\begin{equation}
    \kZ_{\ell m p q}^{\infty,H} = \sum_ 
    {i= 0} ^{2+k}\ \sum_ 
    {j = 0} ^{2+k-i} \kA^{(i,j)}_{\ell m p q} \frac{d^i R^{H,\infty}_{\ell m \omega}}{dr^i}\frac{d^j \bar S_{\ell m \omega}}{d\theta^j} \, , 
    \label{eq:amplitudes_k}
\end{equation} 
for $k\in \{0,1,2\}$, whereas the general form for $k=3$ is identical to $k=0$, as both contain no covariant derivatives in the source term (see Eqs.~\eqref{eq:canonical_EM_tensor_0} and~\eqref{eq:canonical_EM_tensor_3}). This quantity is evaluated at the outer orbit of the SB, $r = r_0$ $\theta = \pi/2$, and for $\omega=\omega_{mpq}$. The $\mathcal{A}$ depend only on the outer and inner orbital parameters of the SB. Next, we see how to recover Eq.~\eqref{eq:amplitude_inf_0_integrated} for $k\in\{1,2\}$.

%
%
\noindent{\emph{$\overset{1}{T} \hspace{0 em} ^{\mu \nu} $: one covariant derivative}} \label{sec:source_1}
%
To obtain the amplitude associated to the dipole term~\eqref{eq:canonical_EM_tensor_1}, we first perform an integration by parts and obtain Eq.~\eqref{eq:amplitude_inf_0} for $k=1$. However, now there is still a covariant derivative applied to $\delta_{(4)}$, so we perform a second integration by parts. This yields a boundary term, which evaluates to zero due to the periodicity of the SB motion, 
and a bulk term where the covariant derivative no longer acts on $\delta_{(4)}$. Since these are covariant derivatives, factors of $\sqrt{-g}$ appear in performing the integration by parts. The result is 
\begin{widetext}
    \begin{equation}
    \kZZ^{\infty,H} = \int_{\mathcal{M}} \int_\mathbb{R} \sqrt{-g} \, \delta_{(4)} 
    \left(S^{\rho (\mu} u^{\nu)} \left(2 \, n_\mu \nabla_\rho n_\nu + n_\mu n_\nu \nabla_\rho 
    \right) \frac{\mD_{nn} ^\ast}{\sqrt{-g}} + (\cdot\cdot\cdot) \right) {\frak F}^{\infty,H} \, e^{i(\omega t^\prime - m\phi^\prime) } \ d\tau \, d^4 x^\prime  \, ,\label{eq:amplitude_inf_1} 
\end{equation}
\end{widetext}
where $(\cdot\cdot\cdot)$ denotes the terms in $n\bar m$ and $\bar m \bar m$. With no derivatives of $\delta_{(4)}$, the integration is easily performed and the result is simply Eq.~\eqref{eq:amplitude_inf_0_integrated} with $k=1$. We find that only three frequencies are excited per azimuthal mode number $m$ at this order: $(m,0,0)$ and $(m,\pm1,0)$. 
%

%
\noindent{\emph{$\overset{2}{T} \hspace{0 em} ^{\mu \nu} $: two covariant derivatives}} \label{sec:source_2}
%
Here the procedure is almost identical to that for one covariant derivative, except that now we have to perform an extra integration by parts. After doing that, we recover the form of Eq.~\eqref{eq:amplitude_inf_0_integrated} with $k=2$. All possible fifteen frequencies are excited by the dynamic quadrupole. Thus, we conclude that the total amplitude of Eq.~\eqref{eq:amplitude_inf} can be written as
\begin{equation}
    Z^{\infty,H}_{\omega \ell m} = \sum_{\substack{p,q}} Z^{\infty,H}_{\ell m p q}  \delta_{(m p q)} \, , \quad Z^{\infty,H}_{\ell m p q}= \sum_k \kZ^{\infty,H}_{\ell m p q} \, .\label{eq:amplitude_total}
\end{equation}
Eq.~\eqref{eq:amplitudes} is now obtained by substituting Eq.~\eqref{eq:amplitudes_k} above. 
%

%
\noindent{\emph{Reflection symmetry: $m\to-m \, , \, \omega\to-\omega$}} \label{sec:ref_sym}
%
The b-EMRI system in the equatorial plane exhibits a symmetry under the simultaneous transformations $t \to -t$ and $\phi\to -\phi$. This symmetry translates into a symmetry in the amplitudes under the simultaneous transformation $\omega\to-\omega$ and $m\to -m$. This symmetry manifests itself as
\begin{equation}
    Z_{\ell(-m)(-p)(-q)} = (-1)^{\ell+p} \overbar{Z}_{\ell mpq} \label{eq:ref_sym}
\end{equation}
For $p=0$, this reduces to the well-known symmetry of the spinning secondary EMRI amplitudes~\cite{Drasco:2005kz,Cardoso:2019nis,Skoupy:2023lih}. 
For that reason, we can say that the b-EMRI system excites a total of eight families of frequencies, each labeled by the vales $p$ and $q$. As an example, the $(m,2,2)$ and $(-m,-2,-2)\equiv(m,-2,-2)$ modes are in the same family. This symmetry is used throughout this work to reduce computational runtime. 
%
%

%
\noindent{\emph{The spin-aligned b-EMRI}} \label{sec:spin_aligned}
%
We will focus mainly on the spin-aligned b-EMRI, when the spin of the SB is aligned with the orbital angular momentum of its motion around the SMBH. This corresponds to setting $\iota_{\rm SB}=0$ in Eqs.~\eqref{eq:SAF1}--\eqref{eq:SAF3}. As we will see, this configuration is simpler, both physically and in terms of implementation.

In Eq.~\eqref{eq:frequencies} we presented a general formula for the frequencies excited by the b-EMRI, and saw that these can be identified with a triplet of integers $(m,p,q)$. In the spin-aligned configuration, the picture is substantially simpler: all terms excite the $(m,0,0)$ mode, and the quadrupole terms also excite the two $(m,\pm2,\mp2)$ modes. Instead of eight, we get only two families of frequencies.

We can easily understand why the situation is so simple in the spin-aligned case. The fact that all terms excite the $(m,0,0)$ mode is not surprising, since all of them correspond to some energy distribution in orbit around the SMBH. The dipole order term does not introduce any extra frequencies because the spin vector does not precess, so the $(m,\pm1,0)$ modes are not excited. Finally, all the complicated frequencies excited at quadrupole order degenerate to the pair $(m,\pm2,\mp2)$. This happens because an observer at infinity cannot distinguish ${\rm \Omega_P}$ and ${\rm \Omega_{SB}}$, since both of them are constant frequencies around the axis defined by the spin $S^\mu$. 
%
\subsubsection{Waveforms and energy fluxes} \label{sec:gws_from_psi}
%
Replacing Eq.~\eqref{eq:amplitude_inf_0_integrated} in Eq.~\eqref{eq:fourier_harmonic} and performing the integral in $\omega$ yields Eq.~\eqref{eq:psi4}.

Besides the strain~\eqref{eq:strain}, energy fluxes can also be calculated from the amplitudes using the results in Ref~\cite{Teukolsky:1974yv}. The fluxes at infinity and on the horizon are, respectively, 
\begin{align}
    \dot{E}^\infty =& \sum_{\ell, m, p, q} \dot{E}^\infty _{\ell m pq} = \sum_{\ell, m, p, q} \frac{|Z^\infty_{\ell m pq} |^2}{4 \pi\omega_{m p q}^2} \, , \label{eq:energy_inf} \\ 
    \dot{E}^H =& \sum_{\ell, m, p, q} \dot{E}^H _{\ell m pq} = \sum_{\ell, m, p, q} \alpha_{\ell mpq}\frac{|Z^H_{\ell m pq} |^2}{4 \pi\omega_{m p q}^2} \, , \label{eq:energy_hor}
\end{align}
where the specific expression for $\alpha_{\ell m p q}$ can be found in Eq.~(4.44) of Ref~\cite{Teukolsky:1974yv}. The expressions for the amplitudes $Z^H$ are functionally the same as for the amplitudes at infinity, except that the homogeneous solution of the Teukolsky equation inside the integral in Eq.~\eqref{eq:amplitude_inf} is $R^\infty$ instead of $R^H$.

\subsection{Weak-field limit of Dixon's moments} \label{app:Newtonian_quadrupole}
%
%
%
In standard textbook presentations of GW generation by a binary system in the weak-field, slow-motion limit~\cite{Poisson_Will_2014, Wald:1984rg}, the source term is the second time derivative of the mass quadrupole moment. We will now recover this result with our formalism.  

Consider the center of mass of the SB at rest in flat spacetime. In that case, we only have two frames: the spin-aligned frame~\eqref{eq:SAF1}-~\eqref{eq:SAF3}, which now plays the role of a static frame with the origin at the SB center of mass, and the binary-aligned frame defined in Eqs.~\eqref{eq:BAF1}-\eqref{eq:BAF2}. There is no distinction between proper time and coordinate time. We start by considering the following decomposition of the quadrupole moment tensor~\cite{Ehlers:1977gyn}:
\begin{equation}
    J^{\mu \nu \rho \sigma} = \Sigma^{\mu \nu \rho \sigma} - u^{[\mu}\Pi^{\nu] \rho \sigma} - u^{[\rho}\Pi^{\sigma] \mu \nu} - 3 u^{[\mu}Q^{\nu][\rho}u^{\sigma]} \, ,
\end{equation}
where $\Sigma$, $\Pi$ and $Q$ are the stress, momentum and mass quadrupole moments, respectively, and $u$ is the four-velocity. In the binary-aligned frame used to obtain Eq.~\eqref{eq:multipoles_explicit}, the mass quadrupole moment has components
\begin{equation}
    Q^{\hat\mu \hat\nu} = \frac{1}{2 } \mu d^2 \delta^{\hat \mu}_2 \delta^{\hat \nu}_2 \, .
\end{equation}
Using Eqs.~\eqref{eq:BAF1}-\eqref{eq:BAF2}, we obtain the components in the static frame:
\begin{align}
    Q^{\mu \nu}(t) = \frac{1}{2 } \mu d^2 \Big(&\sin^2 (\Omega_{\rm SB} t) \delta^{\mu}_1 \delta^{\nu}_1 \\
    & + 2 \sin (\Omega_{\rm  SB} t) \cos (\Omega_{\rm  SB} t) \delta^{(\mu}_1 \delta^{\nu)}_2 \nonumber  \\
    &+ \cos^2 (\Omega_{\rm  SB} t) \delta^{\mu}_2 \delta^{\nu}_2 \Big) \nonumber \, .
\end{align}
In the slow-motion limit, the GWs generated in the far zone, at a distance $R\gg d$, are given, in the Lorenz gauge, by 
\begin{equation}
	\bar{h}_{\mu \nu} (t,\vec{x}) = 4 \int_{\mathbb{R}^3} d^3 x^\prime \frac{T_{\mu \nu}(t-R , \vec{x}^\prime)}{R}\, , 
\end{equation} 
where $R=|\vec{x}-\vec{x}^\prime|$. Replacing the canonical form of the energy momentum tensor~\eqref{eq:canonical_EM_tensor}, we get, at leading order 
\begin{equation}
	\bar{h}^{\mu \nu} (t,\vec{x}) = 2 \frac{ \ddot{Q}^{\mu \nu}}{R}\, , 
\end{equation}
where we used dots to denote differentiation. This is the standard Newtonian quadrupole formula, present, for example, in Eq.~(11.55) of Ref.~\cite{Poisson_Will_2014}. 
%
%
\subsection{Formal adjoint operators $\mD_{\bullet \bullet}^\ast$} \label{app:formal_adjoints}
Here we want to explicitly show how to get from Eqs.~\eqref{eq:T4}--\eqref{eq:amplitude_inf} to Eq.~\eqref{eq:amplitude_inf_0}. Without loss of generality (because of the linearity of the equations), let us look only at one of the terms in ${T} \hspace{0 em}_{4}$, say the term $\mD_{nn} {T} \hspace{0 em}_{nn}$. When replaced in Eq.~\eqref{eq:amplitude_inf}, it yields a term proportional to the integral  
\begin{equation}
   \int_{\mathcal{M}} {\frak F} \, e^{i(\omega t^\prime - m\phi^\prime) } \left(\mD_{nn}  \overset{k}{T} \hspace{0 em}_{nn}\right) d^4 x^\prime \, . \label{eq:A1}
\end{equation}
Before obtaining $\mD_{nn}^\ast$, let us first consider defining the simpler linear differential operator 
\begin{equation}
    \mD = \left(\bar m^\mu \partial_\mu + f \right) \, ,
\end{equation}
where $f$ is a smooth function, and let us look at the integral 
\begin{equation}
   \int_{\mathcal{M}} {\frak F} \, e^{i(\omega t^\prime - m\phi^\prime) } \left(\mD\,  \overset{k}{T} \hspace{0 em}_{nn}\right) d^4 x^\prime \, . 
   \label{eq:A3}
\end{equation}
Performing an integration by parts and applying the divergence theorem yields
\begin{align}
   & \int_{\partial \mathcal{M}} \left( {\frak F} \, e^{i(\omega t^\prime - m\phi^\prime) } \overset{k}{T} \hspace{0 em}_{nn} \right) \bar m^\mu dS_\mu \\ 
   &- \int_{\mathcal{M}} \overset{k}{T} \hspace{0 em}_{nn} \left(\bar m^\mu \partial_\mu + \partial_\mu \bar m^\mu - f \right)  \left({\frak F}\,e^{i(\omega t^\prime - m\phi^\prime) }\right) d^4 x^\prime \, , \nonumber
\end{align}
where $dS_\mu$ is the induced volume form on the boundary $\partial \mathcal{M}$.\footnote{Note that here by integral on the boundary of $\mathcal{M}$ one should understand the limit of integrals on the boundary of bounded domains in $\mathcal{M}$ when these approach the whole of $\mathcal{M}$.} Since the energy momentum tensor is that of a particle eternally in circular orbit, its value at the boundary is not zero. However, since the motion is periodic, the flux at future and past timelike infinity cancel each other out. Thus, the boundary term evaluates to zero, so that \eqref{eq:A3} is equal to 
\begin{gather}
    \int_{\mathcal{M}} \overset{0}{T} \hspace{0 em}_{nn} \left(\mD^\ast \left({\frak F} \, e^{i(\omega t^\prime - m\phi^\prime) }\, \right) \right) d^4 x^\prime \, , \\ \nonumber \\
    \mD^\ast = \left(- \bar m^\mu \partial_\mu - \partial_\mu \bar m^\mu  + f \right) \, .
\end{gather}
Now consider instead a second-order operator such as
\begin{equation}
    \mD_{nn} = -\left(\bar m^\mu \partial_\mu + f_1 \right)\left(\bar m^\mu \partial_\mu + f_2 \right) \, ,
    \label{eq:A7}
\end{equation}
where $f_{1,2}$ are again just smooth functions (actually they are linear combinations of the spin coefficients). Comparing with the result for the formal adjoint of the first-order operator, it easy to check that \eqref{eq:A1} is equal to 
\begin{gather}
   \int_{\mathcal{M}} \overset{0}{T} \hspace{0 em}_{nn} \left(\mD_{nn}^\ast \left( 
 {\frak F} \, e^{i(\omega t^\prime - m\phi^\prime) }  \right) \right) d^4 x^\prime \, , \\ \nonumber \\
    \mD_{nn}^\ast = \left( \bar m^\mu \partial_\mu + \partial_\mu \bar m^\mu  - f_2 \right)\left(\bar m^\mu \partial_\mu + \partial_\mu \bar m^\mu  - f_1 \right) \, . 
\end{gather}
As for $\mD_{n \bar m}$ and $\mD_{\bar m \bar m}$, they are all written in the form of Eq.~\eqref{eq:A7}, so from this result their formal adjoints $\mD_{n \bar m}^\ast$ and $\mD_{\bar m \bar m}^\ast$ can be easily obtained. 

These operators are, of course, the outgoing  radiation gauge metric reconstruction operators commonly found in the literature~\cite{Lousto:2002em, vandeMeent:2015lxa}.
%
%
\subsection{Numerical method and convergence}
%
%
We now focus on the numerical method we employed, with an emphasis on the sum over $(\ell,m)$ harmonics discussed in the main text. We also discuss the results of the benchmarking tests performed.
%
\subsubsection{Numerical setup} \label{sec:num_setup}
%
To obtain the waveform produced by a b-EMRI we face two main challenges. The first concerns obtaining the functional form of the amplitudes $Z$ in Eq.~\eqref{eq:amplitudes_SM} (or more generally Eq.~\eqref{eq:amplitudes_k} including the amplitudes on the horizon), which we do analytically. The second, which we solve using semi-analytic methods, consists of calculating the homogeneous solutions to the Teukolsky equation $R^{H,\infty}$ 
 -- we use the Mano-Suzuki-Takasugi method~\cite{Mano:1996vt,Fujita:2004rb} -- and the spheroidal harmonics -- we use Leaver's algorithm~\cite{Leaver:1985ax,Berti:2005gp}. 

Following this method yields a single amplitude $\kZ^{H,\infty}_{\ell m p q}$, and adding over $k$ yields an amplitude $Z^{H,\infty}_{\ell m p q}$. In principle, to obtain the full waveform of the system we have to add up contributions from all possible $p$ and $q$, and, for each combination thereof, sum over enough $(\ell,m)$ harmonics until the sum converges (to some accuracy of our choice discussed after Eq.~\eqref{eq:Cl} below). This sum is especially important in this system because, contrary to a regular EMRI, the GW content is not necessarily mostly in the $\ell=\pm m = 2$ harmonics.\footnote{This is most prominent for the higher frequency modes with $q=\pm2$, which are sourced by the SB's inner motion: they excite many $(\ell, m)$ harmonics, especially when the SB is farther away from the SMBH or when $\Omega_{\rm SB}$ is larger~\cite{Bonetti:2017hnb}. This happens because the spheroidal harmonic basis is best suited for low-frequency signals and/or signals sourced at the origin~\cite{Boyle:2015nqa}.}

In the remainder of this section we focus exclusively on the fiducial simulation used in the main text (see Tab.~\ref{tab:parameters_fiducial}). All results presented are not exclusive to the fiducial setup and were tested for a variety of system parameters (including runs with $\iota_{\rm SB} \neq 0$) yielding qualitatively identical results.
%
%
\subsubsection{Sum over $(\ell,m)$ harmonics} \label{sec:l_sum}
%
\begin{figure*}[!t]
    \centering
    \includegraphics[width= .48 \textwidth]{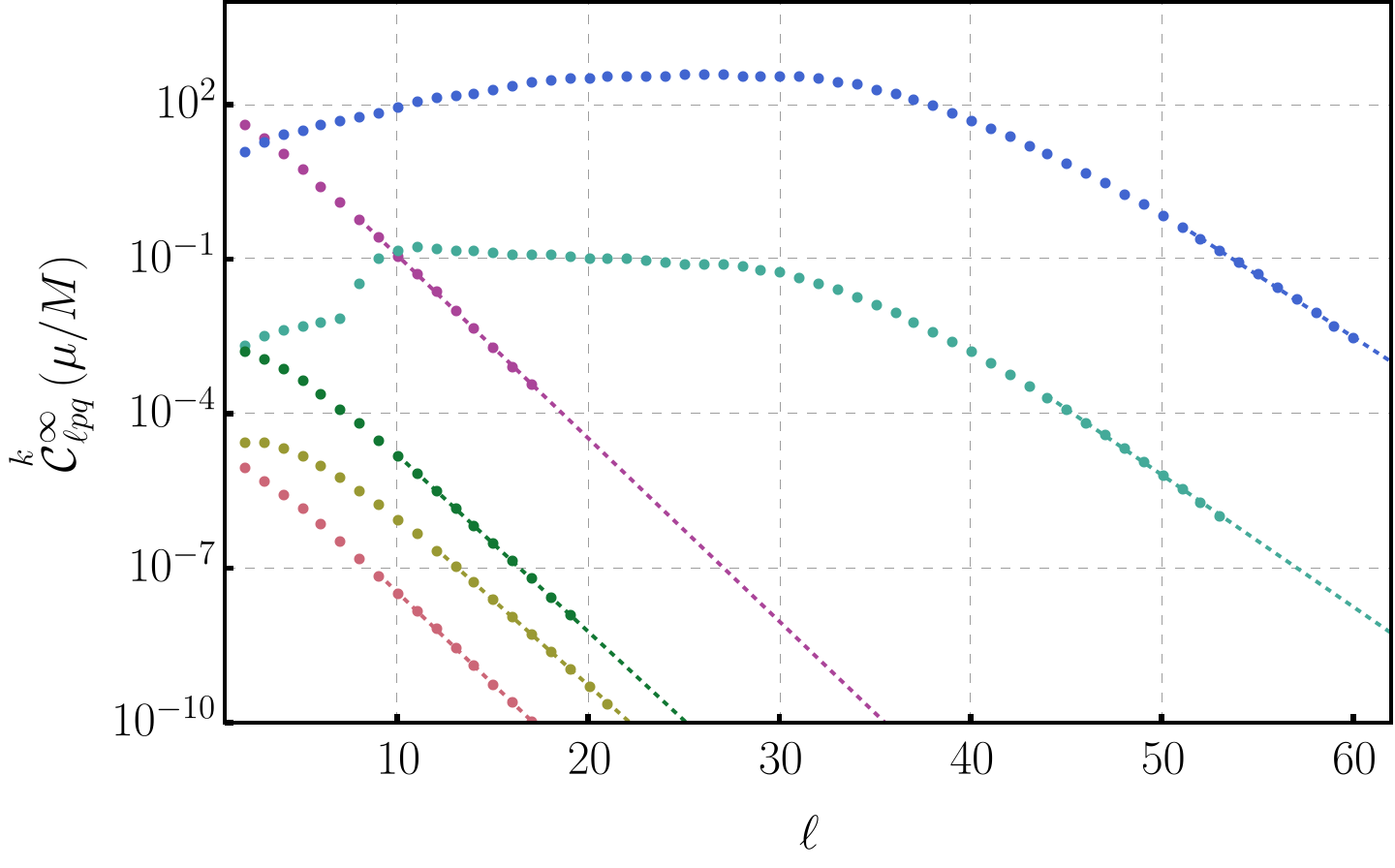}
    \includegraphics[width= .48\textwidth]{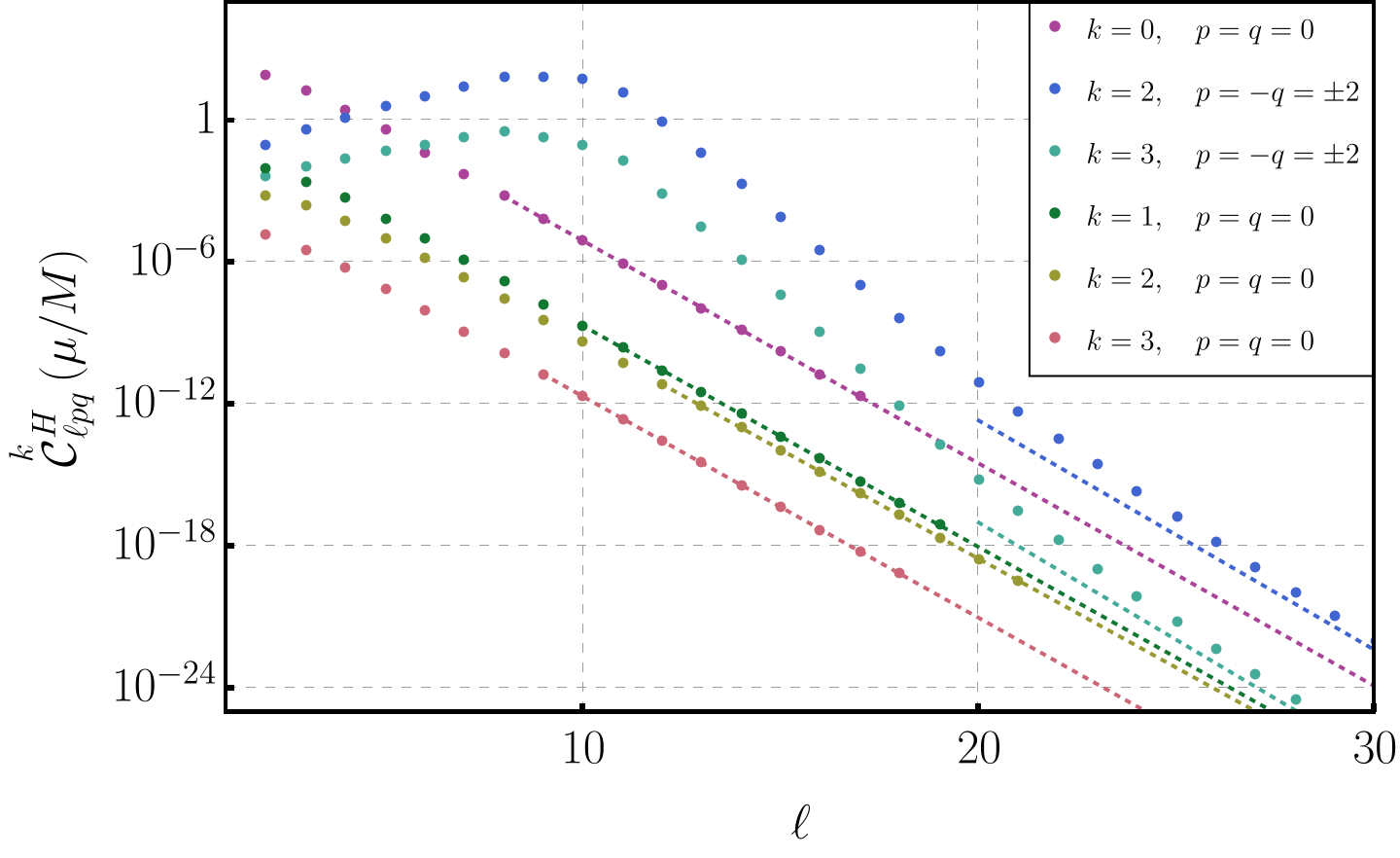}
   \caption{\justifying  The $\mathcal{C}^{H,\infty}_{\ell p q}$ coefficients of Eq.~\eqref{eq:Cl} for increasing $\ell$, for all the families of frequencies (labeled by $p$ and $q$) excited by the spin-aligned b-EMRI, discriminating contributions from the different terms in the energy momentum tensor (labeled by $k$); The parameters of the system are shown in Table~\ref{tab:parameters_fiducial}.
   {\bf Left}: Amplitudes at infinity, $\mathcal{C}^\infty_{\ell p q}$. 
   {\bf Right}: Amplitudes at the SMBH horizon, $\mathcal{C}^H_{\ell p q}$. Each family of frequencies is represented by the values of $p$ and $q$ in the triplet $(m,p,q)$, corresponding to the frequency $\omega_{mpq}$ given in Eq.~\eqref{eq:frequencies}. The term in the energy momentum tensor is indicated by the value of $k\in\{0,1,2,3\}$, Eqs.~\eqref{eq:canonical_EM_tensor_0}--\eqref{eq:canonical_EM_tensor_3}. Dashed lines are fits of the form $C_{\ell p q} \sim 10^{-\beta \ell}$, using the last ten points in each data set. These fits show exponential convergence and, from top to bottom (as presented in the legend), the values of $\beta$ for $\mathcal{C}^\infty_{\ell p q}$ are $\beta=\{0.36, \,0.24,\,0.26, \,0.34,\, 0.33,\, 0.35\}$, while for $\mathcal{C}^H_{\ell p q}$ they are $\beta=\{0.94,\,0.97, \,0.99,  \,0.94,\, 0.92,  \,0.94\}$.}
    \label{fig:convergence_fiducial}
\end{figure*}
We now study the convergence properties of summing over the harmonics. For each value of $k$, $p$ and $q$, we want to determine how many $(\ell,m)$ harmonics have to be included for the waveform to converge. We start with $\ell=2$ and increase $\ell$ one by one, calculating the amplitude for $m\in\{-\ell, ...,\ell\}$. We quantify how large the amplitudes are at a given value of $\ell$ through the quantity 
\begin{equation}
    \kC_{\ell p q} ^{H,\infty} = \sum_{m=-\ell} ^\ell \big|\kZ_{\ell m p q} ^{H,\infty}\big| \, .
    \label{eq:Cl}
\end{equation}
We do this for all harmonics until a convergence criterion is met for $\ell = \ell_{\rm max}$, which should quantify the statement that the amplitudes are sufficiently small for the waveform to have converged. Since we are interested in obtaining the waveform at infinity, we applied the convergence criterion to the amplitudes at infinity. Still, in all simulations, if it were instead applied to the horizon amplitudes it would have been met at the same $\ell$ or earlier. The criterion used was:
\begin{enumerate}
    \item $\kC^\infty_{\ell p q}$ is decreasing for at least the last ten consecutive values of $\ell$;
    \item The last value satisfies $\kC^\infty_{\ell p q} < \max\limits_{\ell^\prime \leq \ell} \ \kC^\infty_{\ell^\prime p q}\times 10^{-5}$.
\end{enumerate}
In Fig.~\ref{fig:convergence_fiducial} we present the values of $\kC_{\ell p q}$ for increasing $\ell$ for all the modes present in the spin-aligned case. We also show, in dashed lines, fits of the form $\mathcal{C}\sim 10^{-\beta \ell} $ obtained using the last ten points in each data set. We find that all modes converge under the criterion described above and that, moreover, for sufficiently large $\ell$ we always get exponential convergence, with $0<\beta \lesssim 1$. 

For the low-frequency modes with $p=q=0$, it is apparent that the only term producing non-negligible amplitudes is the monopole ($k=0$, corresponding to Eq.~\eqref{eq:canonical_EM_tensor_0}), for which convergence was achieved at $\ell=17$. Conversely, for the high-frequency modes with $p=-q=\pm2$, the only non-negligible contribution comes from the dynamic quadrupole ($k=2$, corresponding to Eq.~\eqref{eq:canonical_EM_tensor_2}), for which the sum only converged at $\ell=60$. As for the amplitudes at the horizon, we only show the first 30 values of $\ell$, since the onset of exponential convergence occurs sooner and for higher values of $\beta$. 

A clear feature in Fig.~\ref{fig:convergence_fiducial} are resonances in the high-frequency modes around $\ell=10$. These are present in the amplitudes at the horizon and infinity. These are resonances with QNMs of the SMBH, we expand on this in~\ref{sec:QNMs}.

Based on these results and all the other simulations we ran but don't show, we conclude the following: \emph{The modes excited by the b-EMRI can be separated in two sets: low-frequency, with $q=0$, and high-frequency, with $q=\pm2$.} The low-frequency content is always predominantly in the $(m,p,q)=(m,0,0)$ modes, as sourced by the monopole term, Eq.~\eqref{eq:canonical_EM_tensor_0}. The high-frequency content is distributed in all triplets $(m,p,q)$, with $q=\pm2$, and is predominately sourced by the dynamic quadrupole, Eq.~\eqref{eq:canonical_EM_tensor_2}. Moreover, we find that for $q=\pm2$ the sum over harmonics converges for $\ell = \ell_{\rm max} \sim  r_0 \Omega_{\rm SB}$, with a coefficient $\mathcal{O}(1)$.
%
%
\subsubsection{Benchmarks} \label{sec:benchmarks}
%
Given the complicated nature of the problem, there are two important benchmarks to be considered: the EMRI and the Newtonian limits. We are able to recover standard literature results in both limits, thus strengthening the reliability of the model.
%

\noindent{\emph{The EMRI limit}} \label{sec:EMRI}
%
The first limit we consider is taking $d\to 0$ while keeping ${\rm \Omega_{SB}}$ constant; in this limit, the dipole and quadrupole order amplitudes vanish, leaving only the monopole with $k=p=q=0$. We call this the EMRI limit, as we must recover the amplitudes of an EMRI for a point mass secondary with total mass $2 \mu$. 

In order to compare with known results for EMRIs, let us define the cumulative energy flux up to a given $\ell$, for a pair of $p$ and $q$, as the result of adding up the energy fluxes in all harmonics up to that $\ell$, that is
\begin{equation}
    \dot{E}^{H,\infty} _{\ell pq} = \sum_{\ell^\prime = 2}^\ell \sum_{m^\prime = - \ell^\prime} ^{\ell^\prime} \dot{E}^{H,\infty}_{\ell^\prime m^\prime p q} \, . \label{eq:cumulative_flux}
\end{equation}
We start by computing the cumulative energy flux for modes with $p=q=0$ as a function of $\ell$ for our fiducial simulation. Recall from the previous subsection that the contributions from $k=1,2,3$ are negligible to the low-frequency content, so this energy flux is exclusively due to the monopole amplitudes with $k=0$. We compare our numerical results with the cumulative energy flux calculated using the Black Hole Perturbation Toolkit (BHPT)~\cite{BHPToolkit}. We find that our results agree with the BHPT with a relative precision $<10^{-5}$. The individual amplitudes $Z_{\ell m p q}$ also agree with the BHPT result to a similar precision. 

%
\noindent{\emph{The Newtonian limit.}} \label{sec:newtonian}
%
The second interesting limit is taking the orbital radius of the outer orbit, $r_0$, to be very large. In this limit, the SB is almost at rest in an approximately flat spacetime. Thus, the energy flux is almost exclusively in the high-frequency modes with $q=\pm2$, and is simply given by the Newtonian quadrupole formula~\cite{Peters:1963ux,Poisson_Will_2014}, which, for an equal mass circular binary, takes the form  
\begin{equation}
    \dot{E^\infty} \approx \dot{E}_\text{N} = \frac{2}{5} (2 \mu / d)^5 \, . \label{eq:quadrupole_formula}
\end{equation}
Our aim is to develop a better model that is also valid for finite $r_0$. Since we want an estimate of the energy flux emanating from the b-EMRI's internal motion, we will focus on the high-frequency modes with $q=\pm2$, for which the GW frequency is $\omega \sim \pm 2 \Omega_{\rm SB} $. A naive expectation is that the energy flux (with respect to proper time) measured by a family of observers co-moving with the SB will always be given by Eq.~\eqref{eq:quadrupole_formula}. However, this idea only makes sense if it is possible to define a wave zone around the SB with dimensions $\sim\lambda$ (for $\lambda\sim\pi/{\rm \Omega_{SB}}$ the wavelength of the radiation) satisfying $\lambda\ll\sqrt{r_0^3/M}$, the local radius of curvature. This is precisely the geometric optics limit~\cite{Wald:1984rg}. 

In~\ref{app:model} we develop a model within a geometric optics approximation for how the energy flux detected at infinity is related to the energy flux detected by a family of local observers. The result of this calculation accounts for corrections due to both gravitational and Doppler shift, as well as relativistic beaming. The result is 
\begin{equation}
    \dot{E}_Q = \frac{\dot{E}_N}{(u^t)^2\left(1 - \Omega_0 \mathcal{L}/\mathcal{E}\right)} \, ,
    \label{eq:quadrupole_formula_corrected}
\end{equation}
where $\mathcal{L}= g_{\phi \nu} u^\nu$ and  $\mathcal{E}= - g_{t \nu} u^\nu$ are the energy and angular momentum along $z$ (per mass unit) of the SB. 

\begin{figure}
    \centering
    \includegraphics[width=\linewidth]{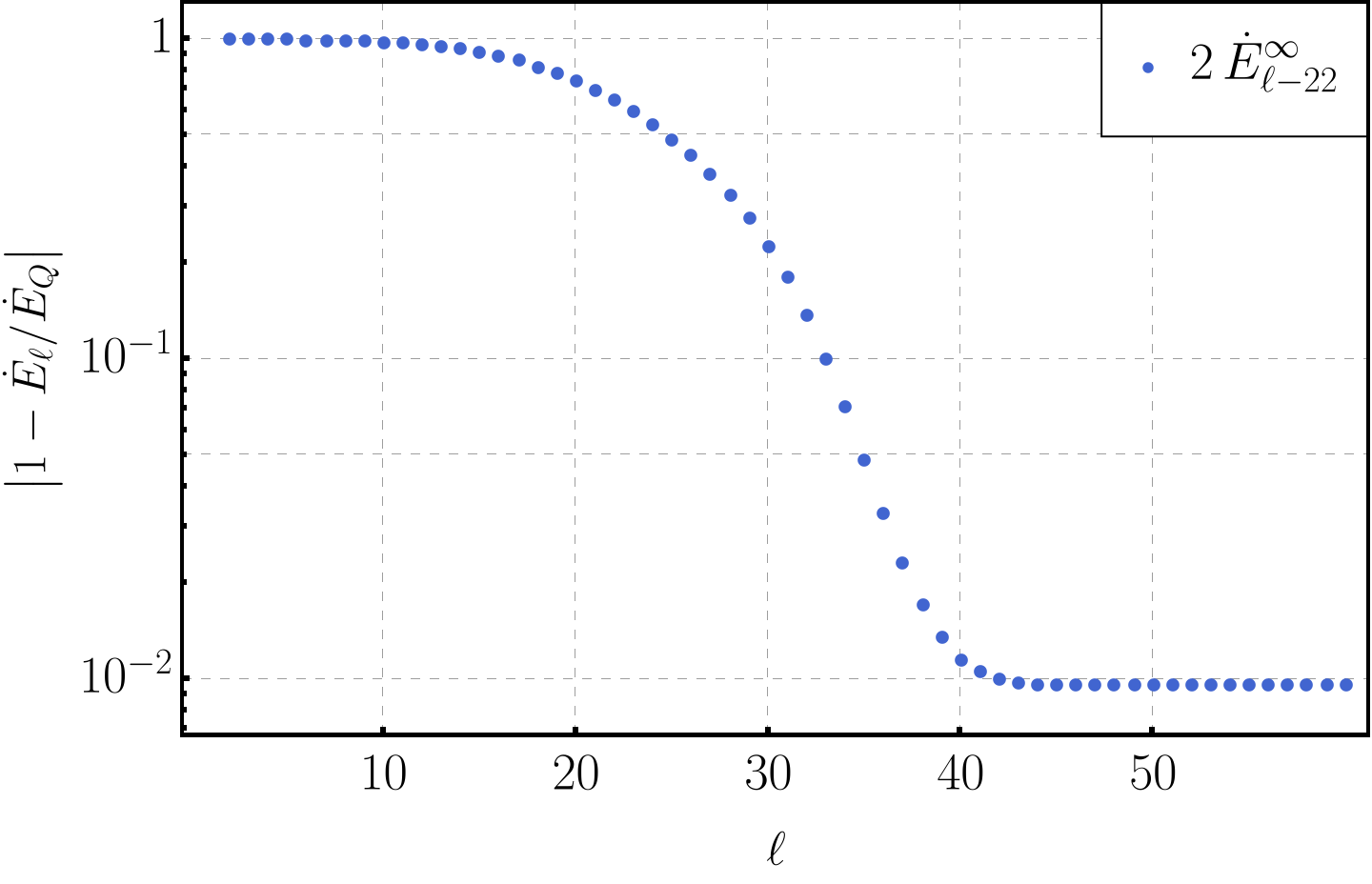}
    \caption{\justifying Cumulative energy flux at infinity~\eqref{eq:cumulative_flux} in modes the modes with $p=-q=\pm2$ (frequencies in Eq.~\eqref{eq:frequencies}), as a function of $\ell$. System parameters are shown in Table~\ref{tab:parameters_fiducial}. Fluxes are normalized to the quadrupole formula for the SB, corrected for relativistic propagation effects in Eq.~\eqref{eq:quadrupole_formula_corrected}. The flux at infinity agrees with the quadrupole formula at a level $<1\%$, even for a system placed at $r_0=10M$.}
    \label{fig:energy_quadrupole}
\end{figure}

In Fig.~\ref{fig:energy_quadrupole} we show the cumulative energy flux at infinity (see Eq.~\eqref{eq:cumulative_flux}) in the $q=\pm2$ modes, normalized to the energy flux predicted by our geometric optics model \eqref{eq:quadrupole_formula_corrected}, for the fiducial simulation. Using the reflection symmetry of the amplitudes (see Eq.~\eqref{eq:ref_sym}), the energy flux in the two $q=\pm2$ modes is just $2 \dot{E_{\ell -2 2}}$. We find agreement between the numerical and analytical results for the flux at infinity to an accuracy $<1 \%$. In this run, the ratio of the wavelength to the local radius of curvature is $\lambda / \sqrt{r_0^3 /M} \sim 10^{-3}$, well within the geometric optics limit. 

%
We tested our formula~\eqref{eq:quadrupole_formula_corrected} in several simulations, and confirmed that it is valid when the parameters of the b-EMRI are within the geometric optics limit. Moreover, it gives a far better prediction than more naive corrections of the Newtonian quadrupole formula. 

We performed simulations with the parameters used to construct Tab. IV of Ref.~\cite{Yin:2024nyz}, and unfortunately were unable to recover their results, neither in the EMRI nor in the Newtonian limits. We understand from the authors of that work that a missing term in the source of the Teukolsky equation and an insufficient number of multipoles in their analysis may explain the disagreement. 
%
\subsection{A relativistic model of Doppler shift and beaming}~\label{app:model}
%
In this section we derive our model for Doppler shift~\eqref{eq:Doppler}, as well as for the modified quadrupole formula~\eqref{eq:quadrupole_formula_corrected}. This derivation is performed under the assumption that it is possible to identify a wave zone around the SB, that is, under the assumption that the wavelength of the radiation, $\lambda\sim\pi/{\rm \Omega_{SB}}$, satisfies $\lambda \ll \sqrt{M/r_0^3}$ (it is much smaller than the local radius of curvature). In that case, we will consider a family of observers co-moving with the SB and located in a normal sphere of radius $\sim\lambda$ around it. The local inertial frame (LIF) of these observers $\{u,e_r,e_\phi,e_z \}$ was defined in Eqs.~\eqref{eq:LIF1}--\eqref{eq:LIF3}. These observers will detect an average energy flux given by Eq.~\eqref{eq:quadrupole_formula}:
\begin{equation}
    \frac{dE_\text{LIF}}{d\tau}  = \dot{E}_N = \frac{2}{5} (2 \mu / d)^5 \, .
    \label{eq:app_LIF_e_flux}
\end{equation}
The approximation we are making implies the geometric optics limit, meaning we can think of the SB as emiting individual gravitons following null geodesics:
\begin{align}
    c:I\subset\mathbb{R} &\to \mathcal{M} \, \\
    s &\mapsto (t(s), r(s), \theta(s), \phi(s)) \, .
\end{align}
If we choose the affine parameter $s$ such that $\dot{c}(s)$ has unit time component in the LIF, then the energy measured for a single graviton by an observer attached to the LIF of the secondary binary is  
\begin{equation}
    \Delta E_\text{LIF} = \hbar \omega = - \hbar \omega \,  g_{\mu \nu} \dot{c}^\mu u^\nu = - \hbar \omega  \left<\dot{c}, u \right>_\text{LIF}\, ,
\end{equation}
where $\hbar$ is the reduced Planck constant and $\omega$ is the angular frequency of the graviton in the LIF. We introduce the notation $\left<\cdot, \cdot\right>_\text{LIF}$ to denote an inner product taken at a particular point in spacetime, in this case at the SB orbit, and write $\Delta E_\text{LIF}$ in this form to emphasize that the graviton's energy-momentum four-vector is $\hbar \omega \dot{c}$. Now consider a family of stationary observers at future null infinity $\mathcal{J}^+$. When the graviton arrives at $\mathcal{J}^+$, the observer at infinity measures its energy to be
\begin{equation}
    \Delta E = - \hbar \omega \left<\dot{c}, \partial_t \right>_{\mathcal{J}^+}\, .
\end{equation}
Substituting the four-velocity of the SB given in Eq.~\eqref{eq:geodesic} yields the ratio of the energies detected by the two observers in the form~\cite{Cisneros:2012sk}
\begin{align}
    \frac{\Delta E}{\Delta E_\text{LIF}} =& \frac{\left<\dot{c}, \partial_t \right>_{\mathcal{J}^+}}{ \left<\dot{c}, u \right>_\text{LIF}} \label{eq:app_ratio} \\
    =& \frac{\left<\dot{c}, \partial_t \right>_{\mathcal{J}^+}}{u^t\left(\left<\dot{c}, \partial_t \right>_\text{LIF} + \Omega_0 \left<\dot{c}, \partial_\phi \right>_\text{LIF} \right)} \nonumber \\
    = & \frac{1}{u^t\left(1+ \Omega_0 \frac{\left<\dot{c}, \partial_\phi \right>_\text{LIF}}{\left<\dot{c}, \partial_t \right>_\text{LIF}} \right) } \nonumber \, , 
\end{align}
where in the last equality we used the fact that $\partial_t$ is a Killing vector field, and so $\left<\dot{c}, \partial_t \right>$ is constant along the null geodesic. Note that this expression is independent of the angular frequency $\omega$.

To obtain the ratio in the denominator, we first write the components of $\dot{c}$ in the LIF $\{u,e_r,e_\phi,e_z \}$ and introduce spherical coordinates $(\vartheta ,\varphi)$ on the sphere of observers attached to this frame:
\begin{align}
    &\dot{c} = (1, \vec{n}) \, , \quad \vec{n} = n^1  e_r + n^2 e_\phi + n^3 e_z \, , \\
    &n^1 = \sin \vartheta \ \cos{\varphi}\, , \\
    &n^2 = \sin \vartheta \ \sin{\varphi} \, ,\\ 
    &n^3 = \cos \vartheta \, .
\end{align}
Then we have 
\begin{align}
    & f_1(\vartheta, \varphi) \coloneqq \left<\dot{c}, \partial_\phi \right>_\text{LIF} = u_\phi + \, n^2  \, (e_\phi)_\phi \, , \\
    & f_2(\vartheta, \varphi) \coloneqq \left<\dot{c}, \partial_t \right>_\text{LIF} = u_t + \, n^2  \, (e_\phi)_t \, .
\end{align}
We conclude that the change in the energy of the graviton depends on the angle it makes with the velocity of the SB (as measured by a local observer). This is precisely the combined effect of Doppler shift and beaming. 

To obtain Eq.~\eqref{eq:Doppler}, note that the frequency of the graviton measured by the LIF observer is $\omega = 2 \omega_{SB}$ (recall Eq.~\eqref{eq:OmegaSB}), as for them the SB is identical to an isolated Newtonian equal mass circular binary in flat space~\cite{Peters:1963ux}. Then, to obtain the maximum blue- and redshifted frequencies $\omega_\pm$, we must maximize and minimize the RHS of Eq.~\eqref{eq:app_ratio}. It is easy to see (and intuitive) that this happens when the graviton is emitted parallel to the direction of motion of the SB so $\vec{n}= (0,\pm1,0)$, yielding
\begin{equation}
    \frac{\hbar \, \omega_{\pm}}{\hbar \, \omega_{\rm SB}} =  \frac{1}{u^t} \left(1 + \Omega_0 \frac{u_\phi \pm (e_\phi)_\phi}{u_t \pm (e_\phi)_t}\right)^{-1} \, .
\end{equation}
A simple algebraic manipulation then gives Eq.~\eqref{eq:Doppler}. This result is not identical to Ref~\cite{Cisneros:2012sk} for the redshifted frequencies; this difference will be explored in future work.

To obtain Eq.~\eqref{eq:quadrupole_formula_corrected} we must assign an angular distribution to the radiated energy. Let us assume that the SB emits, on average, one graviton per proper time interval $\Delta \tau$ with direction following a probability distribution with density function $p(\vartheta, \varphi)$. Then, the average energy of a graviton arriving at $\mathcal{J}^+$ will be
\begin{equation}
    \Delta E = \int_{\mathbb{S}^2} \frac{\Delta E_\text{LIF}}{u^t\left(1+ \Omega_0 \frac{f_1(\vartheta,\varphi)}{f_2(\vartheta,\varphi)} \right) } p(\vartheta,\varphi) \sin \vartheta \, d\vartheta d\varphi \, . 
\end{equation}
%
%
%
%
%
Finally, if one graviton is emitted per interval $\Delta \tau$, then one arrives at $\mathcal{J}^+$ per time interval $\Delta t = u^t \Delta \tau $. Thus the average energy flux at infinity is
\begin{equation}
   \frac{dE}{dt}  = \int_{\mathbb{S}^2} \frac{dE_\text{LIF}/ d\tau}{(u^t)^2\left(1 + \Omega_0 \frac{f_1(\vartheta,\varphi)}{f_2(\vartheta,\varphi)} \right) } p(\vartheta,\varphi) \sin \vartheta \, d\vartheta d\varphi \, , \label{eq:app_almost}
\end{equation}
where $dE_\text{LIF}/ d\tau$ is the average energy flux detected by the observers in the LIF and is given in Eq.~\eqref{eq:app_LIF_e_flux}. Only one task remains, choosing an expression for $p(\vartheta,\varphi)$. This probability density function represents the angular distribution of radiated gravitons, so it must follow the differential energy flux of gravitational radiation for a Newtonian equal mass circular binary, which was calculated in Ref.~\cite{Peters:1963ux}. Thus, if the spin of the SB is aligned with the $e_z$ direction (spin-aligned b-EMRI), we have 
\begin{equation}
    p(\vartheta,\varphi) \propto (1 + 6 \cos^2 \vartheta + \cos^4 \vartheta) \, .
\end{equation}
%
Now, since this density function does not depend on $\varphi$, the integral in $\varphi$ is proportional to
\begin{equation}
   \int_{0}^{2\pi} \frac{u_t + \, n^2  \, (e_\phi)_t}{\left(u_t + \Omega_0 u_\phi\right) + n^2 \left((e_\phi)_t + \Omega_0 (e_\phi)_\phi\right)} \, d\varphi \, ,
\end{equation}
and we have
\begin{equation}
   \left<u, e_\phi \right> = 0 \Leftrightarrow u^t (e_\phi)_t + u^t \Omega_0 (e_\phi)_\phi = 0 \, ,
\end{equation}
so that there is actually no $\varphi$ dependence on the denominator. Therefore, all the terms in the integrand of Eq.~\eqref{eq:app_almost} proportional to $n^2 \propto \sin \varphi$ will vanish, meaning that we only get the constant terms, and so the result is independent of the probability density function. In the end, the resulting expression is remarkably simple:
\begin{equation}
    \frac{dE}{dt} = \frac{dE_\text{LIF}/ d\tau }{(u^t)^2\left(1 + \Omega_0 u_\phi / u_t\right)} \, .
\end{equation}
This may be recast in terms of the energy and angular momentum of the SB to yield the form presented in Eq.~\eqref{eq:quadrupole_formula_corrected}. Finally, in the case of a precessing SB the result is still given by the equation above, although it is not as easy to see: now the terms involving $n^2$ will vanish not only because of the integral over $\varphi$, but also because of the averaging over the precession angle ${\rm \Omega_P} t$. 
%
%
\subsection{Comparison with hotspot data}
%
We now discuss the method we used to obtain the second panel in Fig.~\ref{fig:waveform}. The blue curve corresponds to the instantaneous magnitude of the radiation from our fiducial b-EMRI (see Tab.~\ref{tab:parameters_fiducial}), as detected by the edge-on observer who measures the waveform in the top panel of the same figure. The magnitude is obtained from the intensity of the radiation $I$. For a generic waveform with $+$ and $\times$ polarized components, the instantaneous intensity $I(t)$ is simply~\cite{Poisson_Will_2014}
\begin{equation}
    I(t) = \frac{1}{32 \pi T} \int_{t}^{t+T} \left[ \left( \partial_{t^\prime} h_+\right)^2 + \left( \partial_{t^\prime} h_\times\right)^2  \right] d t^\prime \, , \label{eq:instant_flux}
\end{equation}
where $T$ is to be an integer multiple of the period of the GW. However, in our problem the frequency of the GW changes over time. Taking $T\to \infty$ simply yields Eq.~\eqref{eq:energy_inf}; to capture the energy flux over a small period of time, stable against small variations of $T$, we demand that $2 \pi / \Omega_{\rm SB} \sim T \ll 2 \pi / \Omega_0 $. 

Focusing on the $+$ polarized edge-on waveform, we first calculate $(\partial_t h_+)$, then square it and find the zeros of $(\partial_t h_+)^2$. For each pair of consecutive zeros at $t_1<t_2$, we take $T=t_2-t_1$ and calculate Eq.~\eqref{eq:instant_flux}, associating the result to $t=(t_1+t_2)/2$. This choice allows us to resolve the peak in the second panel of Fig.~\ref{fig:waveform}, despite also leading to a lot of noise when the signal is less regular (in the low amplitude regions). Choosing $T$ such that it encompasses a larger number of zeros leads to a smoother curve, but also compromises the resolution of the peak. 

Finally, the magnitude is obtained in analogy to what is done in the hotspot literature~\cite{Hamaus:2008yw,Rosa:2022toh}
\begin{equation}
    {\rm Magnitude} = 2.5 \log_{10} \left(\frac{I (t)}{{\rm min_{t^\prime} I(t^\prime)}} \right) \, .
\end{equation}
We compared the magnitude of the b-EMRI GW radiation with that of a hotspot -- an isotropic source of electromagnetic radiation -- orbiting a SMBH, with the same orbital and SMBH parameters as in Tab.~\ref{tab:parameters_fiducial}. The hotspot curve (black dashed curve of the second panel of in Fig.~\ref{fig:waveform}) was obtained using the software GYOTO~\cite{Vincent:2011wz}, and corresponds to the magnitude of electromagnetic radiation seen by an edge-on observer. 

There are three main differences between our setup and the hotspot calculation. First, in GYOTO, photons follow null geodesics, contrary to the GWs obtained with our model. Second, for convergence purposes, the hotspot has radius of $0.5 M$, much larger than the dimensions of the SB, $d=0.05 M$. Lastly, the SB does not emit GWs isotopically. These factors likely weigh in on the quantitative differences between the two curves.




A final note regarding the sketches in the top panel of Fig.~\ref{fig:waveform}, which illustrate the position of the SB relative to the SMBH. These sketches actually represent the location of the hotspot centroid in the image plane, as computed by GYOTO. By aligning the peaks in magnitude of the SB and the hotspot, we estimate the SB's position to be roughly that of the hotspot.
%
\subsection{Resonance with SMBH quasi-normal modes} \label{sec:QNMs}
%
Here, we explore the concept of b-EMRIs as ``tuning forks" — systems capable of exciting frequencies that resonate with the natural modes of the SMBH~\cite{Cardoso:2021vjq}. The characteristic, resonant frequencies of BHs are the QNMs referred in the main text~\cite{Berti:2009kk,Berti:2025hly}. These frequencies are complex numbers, due to the leakage of energy through the horizon and to infinity. 

QNM frequencies are characterized by three numbers: $\ell$, $m$ and an overtone number $n$. Generically, the frequencies take values $\left|M \omega^{\rm QNM}_{lmn}\right| \gtrsim 1 $, far too high to be excited by regular EMRIs. In contrast, the high-frequency modes of a b-EMRI can fall right into the QNM range of the SMBH, since $M \Omega_{\rm SB } \sim 1$ or larger, for astrophysically relevant parameters (see Table~\ref{tab:parameters_fiducial}).

Analytical expressions for the QNM frequencies are available in the eikonal limit $\ell \gg 1$~\cite{Cardoso:2008bp,Berti:2009kk,Yang:2012he,Dolan:2010wr}. For the $m=\ell$, $n=0$ mode of a Kerr BH with spin $a$ we have~\cite{Dolan:2010wr}
\begin{align}
    \omega^{\rm QNM}_{m m  0} = \frac{1}{b}\Big(m& + \frac{1-2x}{2\sqrt{1-x^2}}(1-i)+ \frac{1-2 x}{216 m (1-x^2)^2} \label{eq:eikonal_QNMs} \\ 
    & \times(-281 + 44 x -161 x^2)\Big) + \mathcal{O}(m^{-2}) \, ,\nonumber
\end{align}
where $x = a/b$ and 
\begin{align}
    & b = 3\sqrt{M \hat r} - a \, , \\ 
    & \hat{r} = 2 M \left[ 1 + \cos \left( \frac{2}{3} \arccos\left(-a/M\right)\right) \right] \, . 
\end{align}
A similar expression exists for the counterrotating modes $m=-\ell$, but we limit the discussion to $m>0$ using the reflection symmetry of the b-EMRI amplitudes discussed in Sec.~\ref{sec:source_B_EMRI}. The QNMs described by Eq.~\eqref{eq:eikonal_QNMs} are the most long-lived, and are thus expected to be the best candidates for resonant excitation. 

If the frequency of the radiation produced by the b-EMRI matches the real part of the QNM frequency, $\omega_{mpq} \approx \omega_{mm0}$, we can expect this mode of the SMBH to be excited. This would manifest as a peak in the amplitude for $Z^{\infty,H}_{m m p q}$. Taking only the leading term in Eq.~\eqref{eq:eikonal_QNMs}, we can predict this peak to take place for a mode with $m$ given by Eq.~\eqref{eq:resonance_estimator}. Moreover, since we only have high-frequency modes with $q=\pm2$, there will be a single resonance for each value of $q$. 


As we saw in the main text, the frequency of the $m=\ell=8$ QNM of the SMBH is compatible with a feature in the spectrogram in Fig.~\ref{fig:waveform}, as predicted using Eq.~\eqref{eq:resonance_estimator}. For b-EMRIs with other parameters, the resonance will happen for a different mode; in all cases we studied, this mode could be predicted with Eq.~\eqref{eq:resonance_estimator}. 

Further evidence of the b-EMRI resonating with the SMBH can be seen in Fig.~\ref{fig:convergence_fiducial}, where we see peaks in the amplitudes of high frequency $q=\pm2$ modes around $\ell=8$.
Looking at the individual amplitudes $Z^{\infty,H}_{\ell m p q}$, we found the most resonant modes to be $\ell=m\sim 8,9,10$. The frequency of the radiation of the b-EMRI in these modes is given by the triple $(m,p,q)$~\eqref{eq:frequencies} and is 
\beq
&M \omega_{(8,-2,2)} \approx 2.33  \, , \\ 
&M \omega_{(9,-2,2)} \approx 2.36  \, , \\ 
&M \omega_{(10,-2,2)} \approx 2.39  \, .
\eeq
We can calculate the fundamental QNM modes to be~\cite{Leaver:1985ax,Berti:2009kk,istgritlink,ebertilink}
\beq
    &M \omega^{\rm QNM}_{(8,8,0)}& \approx 2.30 - 0.0861 \, i \, , \\ 
    &M \omega^{\rm QNM}_{(9,9,0)}& \approx 2.58 - 0.0863 \, i \, , \\ 
    &M \omega^{\rm QNM}_{(10,10,0)}& \approx 2.87 - 0.0864 \, i \, .
\eeq
Indeed, for $\ell=m=8$, the frequency of the b-EMRI mode is within $1 \%$ of the real part of the corresponding QNM. This is the mode for which we find the highest excitation at the level of horizon amplitudes, as well as the feature in the spectrogram. The excitation of the $\ell=9,10$ is less evident, with the real part of the QNM frequency only within $20 \%$ of the frequency of the b-EMRI radiation.

Thus, b-EMRIs can indeed act as tuning forks, resonantly exciting the modes of the SMBH~\cite{Cardoso:2021vjq,Yin:2024nyz}. Moreover, it is another instance of the modes resonating with the QNMs while having a substantially \emph{smaller} frequency than that of the excited QNM. This topic is still poorly understood and will be the subject of further work.
%
\subsection{Phenomenological models}
%
In this final section, we discuss the approaches used in the literature to model Doppler-modulated signals from binary systems; we also discuss the method used to calculate the mismatch $\mathfrak{M}$ in Fig.~\ref{fig:mismatch}.  
%
\subsubsection{Non-relativistic model of Doppler shift}
%
The typical model used in the literature to study the effects Doppler modulation merely introduces phase shifts in the waveform of an isolated binary, which is often taken to evolve due to GW emission~\cite{Bonvin:2016qxr,Inayoshi:2017hgw,Robson:2018svj,Tamanini:2019usx,Meiron:2016ipr, Wong:2019hsq,Randall:2018lnh,Han:2018hby,Yan:2023pyo,Zwick:2025wkt,Takatsy:2025bfk}. Since such an evolution is not included in our model, we apply the phenomenological model to a circular, equal mass monochromatic binary. Moreover, we will limit ourselves to the case where the spin of the SB is aligned with its orbital angular momentum, $\iota_{\rm SB}=0$. The time domain waveform produced by an isolated circular binary is 
\begin{align}
h_{+} &= \frac{2 \mathcal{M}^{5/3}}{D_{\rm L}} (\pi f_{\rm gw})^{2/3}\left(1+\cos^{2}\theta\right) \cos \varphi_{\rm gw}(t)\, ,\label{eq:phenom_1}\\
h_{\times} &=- \frac{4 \mathcal{M}^{5/3}}{D_{\rm L}} (\pi f_{\rm gw})^{2/3} \cos\theta \sin  \varphi_{\rm gw}(t)\label{eq:phenom_2}\,,
\end{align}
where $\mathcal{M} = 2^{-1/5} \mu$ is the chirp mass of the binary, $D_{\rm L}$ is the luminosity distance, $f_{\rm gw} = 2 \sqrt{2 \mu/d^3}$ is the frequency of the GWs, and $\theta$ is the polar angle between the observer and the spin of the binary. The phase is simply $\varphi_{\rm gw} = 2 \pi f_{\rm gw} \, t + \varphi_0$, where $\Omega_{\rm GW}$ is the angular frequency, $t_0$ is the initial time and $\varphi_0$ is the initial phase.

We now consider this binary to be moving in a circular orbit with radius $r_0$ and angular frequency $\Omega_0$ around a SMBH. The observer will then see the SB with a velocity along its line of sight given by 
\begin{equation}
    v_{\parallel}(t) = v_{\rm los} \cos \left[\Omega_0 (t-t_0)\right] \,  ,
\end{equation}
where $v_{\rm los}$ is the maximum velocity along the line of sight and $t_0$ encodes the position of the SB around the SMBH at $t=0$. The observed frequency $f_{\rm obs}$ will be Doppler-shifted accordingly. In the non-relativistic regime, this yields 
\begin{equation}
    f_{\rm obs} (t) = (1 + v_{\parallel}(t))f_{\rm gw} \, .
\end{equation}
Then, the observed signal~\eqref{eq:phenom_1}-\eqref{eq:phenom_2} will be changed by taking $f_{\rm gw}\to f_{\rm obs}$ and
$$\varphi_{\rm gw}\to \varphi_{\rm obs} = 2 \pi \int_0^t f_{\rm obs} (t^\prime) \, dt^\prime + \varphi_0 \, .$$

This is the non-relativistic model we used to compare with our waveforms. There are, in total, seven independent parameters, $\{\mathcal{M}/D_{\rm L},\, \Omega_0  ,\,  \theta ,\, \varphi_0,\,  t_0 , \, v_{\rm los} ,\, f_{\rm gw} \} \eqqcolon \bm{\lambda}$. Then a phenomenological waveform will be a function
\begin{equation}
    h_{\rm Phen}(t) =  h_{\rm Phen}(\bm \lambda; t) \, .
\end{equation}
One could also consider a similar model for a binary with non-zero inclination with respect to its orbit around the SMBH, which would add one more parameter $\iota_{\rm SB}$.
%
\subsubsection{Calculating the mismatch $\mathfrak{M}$}
%
Consider two distinct time domain waveforms, $h_{1}$ and $h_{2}$. We define their inner product to be~\cite{Owen:1995tm,Apostolatos:1995pj,Flanagan:1997kp}
\begin{equation}
    \left( h_1 | h_2 \right) = 4 \, {\rm Re} \int_{-\infty}^\infty \tilde{h}_1^\ast (f) \tilde{h}_2 (f) \, df  \, ,
\end{equation}
where $\tilde h _{1,2}$ are the Fourier space waveforms. When studying the distinguishability of different waveform models, this quantity is usually weighted by the power spectral density of the detector~\cite{Owen:1995tm,Apostolatos:1995pj,Flanagan:1997kp}. Since our signal is almost monochromatic, we opt to use a flat power spectral density
\footnote{Under this assumption, we establish that the phenomenological models are not able to capture the strong field features in the b-EMRI waveform. Real systems are not monochromatic since the signal chirps as the b-EMRI evolves through GW emission. Moreover, b-EMRIs are candidates for multiband detections where the low and high frequency parts of the signal are observed simultaneously~\cite{LISA:2022kgy,Barausse:2020rsu}. A detailed study of detectability of these systems must account for these effects, both of which require specifying the detector noise curve.}. We can also define the overlap of two waveforms 
\begin{equation}
    \mathcal{O}(h_1,h_2) = \frac{\left(h_1 | h_2 \right)}{\sqrt{\left(h_1 | h_1 \right)\left(h_2 | h_2 \right)}} \, ,
\end{equation}
which measures how similar the waveforms are, weighted by their signal-to-noise ratio (SNR). 

Now consider a waveform obtained with our b-EMRI model, $h_{\rm bE}$. This will be a function of all the parameters of our system, namely the orbital frequency $\Omega_0$ and the $\theta$ coordinate of the observer. We want to find the parameters $\bm{\lambda}$ that maximize the overlap $\mathcal{O}(h_{\rm bE}, h_{\rm Phen}(\bm \lambda))$. In doing so, we choose to fix three parameters in $\bm \lambda$: $\mathcal{M}/D_{\rm L}$, $\theta$ and $\Omega_0$. Fixing $\mathcal{M}/D_{\rm L}$ does not affect the result because $\mathcal{O}$ is weighted by the SNR. We fix $\Omega_0$ to be the same as for the $h_{\rm bE}$; this is justified, as otherwise the period of the Doppler modulation would be different in the two waveforms, which would  lead to a smaller overlap over multiple cycles. We also fix $\theta$ to the same value as for $h_{\rm bE}$, which only affects the relative amplitude of the two polarizations, so it should not affect our results. 

Thus, we define the optimized mismatch between $h_{\rm bE}$ and $h_{\rm Phen}$ as
\begin{equation}
    \mathfrak{M} =1- \max_{\substack{\varphi_0, t_0,\\v_{\rm los},f_{\rm gw}}} \mathcal{O}(h_{\rm bE}, h_{\rm Phen}(\bm \lambda)) \, .
\end{equation}
Note that this quantity is not the commonly used fitting factor~\cite{Flanagan:1997kp,Owen:1995tm,Apostolatos:1995pj}, which implies minimizing over all parameters. Still, it is a good approximation, allowing us to draw conclusions about the validity of using the phenomenological model to describe the waveforms of b-EMRIs in the strong field regime.

The b-EMRI waveforms used for the mismatch calculation and to produce Fig.~\ref{fig:mismatch} were not obtained with the parameters of the fiducial simulation (see Tab.~\ref{tab:parameters_fiducial}). Instead, for the blue curve ($r_0=10 M$) we chose $M\Omega_{\rm SB}\approx 0.65$ so that $\Omega_{\rm SB}-\Omega_{\rm P} = 20 \, \Omega_0$. This makes the b-EMRI exactly periodic with period $2\pi / \Omega_0$, leading to pseudospectral convergence in calculating the Fourier transform of $h_{\rm bE}$. 

For the red (R) and green (G) curves in Fig.~\ref{fig:mismatch}, we fixed $M \, \Omega_{\rm SB}$ to the same value as for the blue curve and chose
\begin{align*}
    R & : \ \Omega_{\rm SB}- \Omega_{\rm P} = 10 \, \Omega_0 \implies r_0 \approx 6.35 \ M \approx 6 \ M\, ,\\ 
    G & : \ \Omega_{\rm SB}- \Omega_{\rm P} = 30 \, \Omega_0 \implies r_0 \approx 13.05 \ M\approx 13 \ M \, .
\end{align*}
All curves were obtained for SMBH spin $a=0.7 \, M$, $\mu = 10^{-4} \, M$ and $\iota_{\rm SB}=0$. 

\end{document}